# AI-based Reconstruction for Fast MRI – A Systematic Review and Meta-analysis

Yutong Chen[1,2,3], Carola-Bibiane Schönlieb[4], Pietro Liò[5], Tim Leiner[6], Pier Luigi Dragotti[7], Ge Wang[8], Daniel Rueckert[9,10], David Firmin[1,2], and Guang Yang[1,2]

[1] National Heart & Lung Institute, Imperial College London, London SW3 6NP, U.K.

[2] Cardiovascular Research Centre, Royal Brompton Hospital, London SW3 6NP, U.K.

[3] Faculty of Biology, University of Cambridge, Cambridge CB2 1RX, U.K..

[4] Department of Applied Mathematics and Theoretical Physics, University of Cambridge, Cambridge CB3 0WA, U.K.

[5] Department of Computer Science and Technology, University of Cambridge, Cambridge CB3 0FD, U.K.

[6] Utrecht University Medical Centre, 3584 CX Utrecht, Netherlands

[7] Department of Electrical and Electronic Engineering, Imperial College London, London SW7 2BU, U.K.

[8] Biomedical Imaging Center, Rensselaer Polytechnic Institute, Troy, NY USA 12180

[9] Department of Computing, Imperial College London, London SW7 2BU, U.K.

[10] Institute for Medical Informatics, Statistics and Epidemiology, Technical University of Munich, 81675 Munich, Germany

# ABSTRACT

Compressed sensing (CS) has been playing a key role in accelerating the magnetic resonance imaging (MRI) acquisition process. With the resurgence of artificial intelligence, deep neural networks and CS algorithms are being integrated to redefine the state of the art of fast MRI. The past several years have witnessed substantial growth in the complexity, diversity, and performance of deep learning-based CS techniques that are dedicated to fast MRI. In this meta-analysis, we systematically review the deep learning-based CS techniques for fast MRI, describe key model designs, highlight breakthroughs, and discuss promising directions. We have also introduced a comprehensive analysis framework and a classification system to assess the pivotal role of deep learning in CS-based acceleration for MRI.

**KEYWORDS** | Compressed sensing; deep learning; neural network; magnetic resonance imaging

# I. INTRODUCTION

Magnetic resonance imaging (MRI) is a state-of-the-art medical imaging technique that can determine structural and functional status of body tissues and organs [1]. However, the prolonged MRI acquisition time [2], [3] increases the scanning cost and limits its use in emergency settings. Moreover, subjects have to lie still in the scanners and even hold their breath for thoracic or abdominal imaging [1]. Hence, the slow acquisition of magnetic resonance (MR) images presents significant inconvenience to patients and healthcare systems alike.

The reason of the slow MR acquisition rate is that unlike other imaging modalities, e.g., X-ray and CT, magnetic resonance (MR) data are acquired in the k-space. The k-space is related to the image domain via Fourier transform [1], [4] and represents the spatial frequency information. During MRI acquisition, measures in the k-space are taken sequentially rather than simultaneously, thus prolonging the scanning time.

To address this limitation, the k-space can be undersampled, i.e., not sampled entirely. The missing k-space data are then inferred from the existing k-space points. This leads to an acceleration inversely proportional to the undersampling ratio. For example, if only 50% of the k-space is sampled, the acceleration is 2-fold (excluding scanning preparation and/or pre-scanning time). Among different undersampling techniques [1], compressed sensing (CS) yields an aggressive acceleration rate up to 12.5 fold [5]. CS assumes that if the undersampled signals can be compressed accurately, they can be decompressed or reconstructed accurately [2], [6] without the need of full sampling. Thus, CS extrapolates unknown k-space signals from existing ones, akin to image super-resolution techniques that increase image resolution by reconstructing high-frequency image details [7].

Driven by the growing research on deep learning in computer vision, deep learning-based algorithms have gained popularity for CS based-MRI reconstruction (CS-MRI). Deep learning techniques utilise artificial neural networks to learn the CS reconstruction process. Compared with traditional non-deep learning-based approaches, deep learning enables higher quality reconstruction [3], [8]–[10] and acceleration ratio of MRI acquisition. With an exponentially increasing interest towards deep learning-based CS-MRI, the complexity and diversity of the reconstruction algorithms have increased dramatically. Motivated by this rapidly expanding field, we have conducted a systematic review and the first meta-analysis to summarise the development of deep learning-based CS-MRI. We will outline the background of deep learning in CS-MRI reconstruction, review each algorithmic category, present the results of meta-analysis, and conclude with an outlook on deep learning-based CS-MRI acceleration.

## A. Deep Learning

The power of deep learning centres on its capacity to model complex input-output relationships with a large number of parameters in an artificial neural network (ANN) [11]. An ANN consists of an input layer of nodes, followed by multiple hidden layers and an output layer (Supplementary Figure 1). In CS-MRI, each input node represents a pixel in the undersampled MR image. The pixels are weighted and summed to form the input for the next layer, after a non-linear activation function [12]. The subsequent hidden layers perform a similar process to produce the final reconstructed image.

The connection weights are network parameters that are optimised such that the outputs from the network are as similar to the target outputs as possible. That is, the network weights need to be tuned to minimise the difference between the fully sampled images and the reconstructed images. This process of weight optimisation is known as training the ANN. The training process is guided by the difference, or error between actual and desired outcomes, described by a loss function. In other words, the network receives the undersampled images and outputs the reconstructed versions. The loss function computes the discrepancy between actual and desired outputs and utilises this information to update the parameters of the ANN that model an optimal CS-MRI reconstruction process.

## B. End-to-End

Deep learning-based CS-MRI models fall into 2 main categories: end-to-end and unrolled optimisation [13]–[15]. An end-to-end technique models the CS-MRI reconstruction process directly. In CS-MRI, the process of acquiring undersampled images is:

$$y = UFx, \tag{1}$$

where $x$ is the fully sampled image, $Fx$ is the Fourier transform of the image, i.e., its k-space representation, $U$ is a binary matrix of zeros and ones that denote which k-space locations to sample, and $y$ is the undersampled k-space data [2]. End-to-end techniques model the inverse acquisition or reconstruction process directly, mapping from $y$ to $x$, and hence the name 'end-to-end'. Because of this direct mapping, the reconstruction process is usually fast [14].

An advantage of end-to-end models is that advances from other fields of deep learning are transferable to end-to-end designs (Supplementary Table 1). For example, U-Net [16], a deep learning model originally developed for image segmentation, was readily applied to reconstruct MR images in an end-to-end manner without major modifications [3], [17]. Similarly, self-attention mechanism—designed to enhance natural image processing—was incorporated to deep end-to-end models with performance improvement over U-Net [18]. One limitation is that end-to-end models tend to require a larger sample size to train [13].

## C. Unrolled Optimisation

Unrolled optimisation combines deep learning with traditional iterative CS algorithms. Traditional CS techniques solve the general problem of image recovery:

$$\frac{1}{2}\left\|UFx - y\right\|_2^2 + R(x), \tag{2}$$

in which $x$ is the reconstructed image and the first term enforces the data fidelity, i.e., the reconstructed image does not differ from the undersampled one at the sampled k-space locations. The second term imposes regularisation, typically sparsity constraints, on the reconstructed image to satisfy the CS criteria. Deep learning networks form the regulariser term (See Table 1 for examples). That is, deep learning models are designed to learn the regularisation methods to constrain image reconstruction, rather than directly modelling the reconstruction process itself. From a Bayesian perspective, the regulariser term represents the prior knowledge about the property of the reconstructed image [19], e.g., sparsity (Supplementary Section II-A). Therefore, compared with end-to-end techniques, unrolled optimisation incorporates prior domain knowledge about the expected property of MR images [15]. This reduces the solution space and facilitates model convergence and performance [13]. It may underpin the superior performance of unrolled optimisation methods compared with end-to-end ones [15] with fewer parameters [14].

Furthermore, different networks with different weights or parameters can be used in different iterations of unrolled optimisation. In each iteration, each subnetwork has a relatively small receptive field and can perform the local transformation. This can avoid overfitting as may occur in end-to-end models [10]. Compared with using the same network with the same weights in each iteration, this no-weight sharing approach has demonstrated superior performance [9], [20], [21] with some exceptions [22]. Building upon this no-weight sharing approach, Zeng et al. incorporated dense connections between subnetworks, a technique inspired from image super-resolution literature. This allows each subnetwork to receive the output from all the preceding subnetworks [23]. These developments show that unrolled optimisation is a robust, flexible and powerful deep learning-based CS-MRI technique.

**Table 1: Summary of the regulariser terms used in deep unrolled optimization and their relationship to the corresponding traditional CS techniques.**

| Traditional | Deep learning-based | Regulariser | References |
|---|---|---|---|
| DL[a] | Deep DL | $\lVert z \rVert_1$ | [24], [25] |
| pFISTA-SENSE[b] | pFISTA-SENSE-Net | $\lVert \Psi x \rVert_1$ | [26] |
| TV[c] | TVINet | $\lVert \nabla x \rVert_1$ | [27] |
| IFR-CS[d] | IFR-Net | $\sum_i \lambda_i \lVert k_i x \rVert$ | [20] |
| SLR[e] | H-DSLR | $\lVert \mathcal{T}(\mathcal{F}\nabla x) \rVert_*$ [f] | [28] |
| SToRM[g] | MODL + SToRM | $\lVert R(x) \rVert^2 +$ $\mathrm{Tr}(x^T L x)$ | [29] |
| fields-of-experts[h] | VN | $\sum_i f_i(k_i x)$ | [30] |

[a]**DL**: Dictionary learning. Here $z$ is the latent dictionary representation or transformation of the input image. The regulariser term enforces sparsity on this dictionary representation to satisfy CS criteria. In deep dictionary learning, multiple layers of dictionaries learn this latent transform $z$.

[b]**pFIST-SENSE**: Projected fast iterative soft-thresholding algorithm-sensitivity encoding. It use a transform $\Psi$ to enforce sparsity of the reconstructed image $x$. In the deep learning version, this transform is replaced with a neural network.

[c]**TV**: Total variation. It enforces smoothness on the reconstructed image by minimising changes in the gradient of the image. In the deep learning version, the gradient operator $\nabla$ is replaced with a neural network.

[d]**IFR-CS**: Iterative feature refinement-compressed sensing. The regulariser applies convolutional filters $k_i$ to the reconstructed image.

[e]**SLR**: Sparse and low rank approach. The regulariser minimises the nuclear norm, i.e., the rank of the Hankel matrix $\mathcal{T}$. In the deep version, this operation is replaced with a neural network.

[f]The * subscript means the nuclear norm.

[g]**SToRM**: Smoothness regularization on manifolds. The regulariser has a general form $R(x)$ which can be replaced with a neural network. The second term is a SToRM prior, which uses a Laplacian manifold $L$ to exploit similarity beyond local neighbour.

[h]**Fields-of-experts**: Convolutional kernels $k_i$ operate on the reconstructed image $x$, followed by trainable activation functions $f_i$.

Compared with end-to-end models, the iterative nature of unrolled optimisation increases the computation time. This arises from the need to update both the network weights and the reconstructed images to maintain k-space data fidelity (2). In contrast, end-to-end methods only need to optimise and update network parameters during the training procedure. To mitigate the iterative nature of updating the reconstructed image, one approach trains the deep learning-based regulariser term alone without the data fidelity term [31]–[34] in (2). Then during image reconstruction, the trained regulariser is re-incorporated to optimise the reconstructed image. This solution decomposes the process of optimising the network parameters and images into optimising them separately.

The second, and more popular approach, is to train the unrolled model in an end-to-end fashion, by expressing (2) in a close form. To illustrate, the deep cascaded convolutional neural network (DC-CNN) [10] applies the following loss function to update the network parameters and reconstructed images:

$$\underset{x,\theta}{\mathrm{argmin}} \left|\left| UFx - y \right|\right|_2^2 + \lambda \left|\left| x - f(x|\theta) \right|\right|, \tag{3}$$

where $\theta$ denotes all network parameters, $f(x|\theta)$ the output of the deep learning regulariser, and $\lambda$ is a scalar to adjust the relative contributions of the two terms. The close form of (3) is:

$$x = \frac{\lambda y + F f(x|\theta)}{\lambda + 1},$$

at k-space locations that are sampled and

$$x = f(x|\theta),$$

at k-space locations that are not sampled.

We can also interpret this closed form as another computation layer, called the *data consistency* layer, in deep learning models. The process of iteratively reconstructing the MR images and updating model parameters is a cascade of alternating model reconstruction and data consistency reinforcement (Supplementary Figure 2). In other words, this cascade of neural networks becomes an end-to-end model and is trained in the same fashion. DC-CNN has since then become an integral part of subsequent model designs [8], [9], [34], [35], [36, p. 2], [37]–[40]. However, DC-CNN is not computationally efficient for parallel imaging-based CS acquisition [41], potentially preventing wider spread of DC-CNN-based methods. Moreover, one cannot always derive the closed form of other loss functions as trivially as in DC-CNN [42]. To circumvent this problem, some models apply simple gradient descent [30], [43]–[45], conjugate gradient descent [22], [46], [47], or auxiliary variables [21] (Supplementary Section II-B to II-D), to implement unrolled optimisation in an end-to-end fashion to facilitate the model training process.

While unrolled methods can be trained end-to-end, end-to-end methods can incorporate features of unrolled optimisation. For example, various end-to-end models integrate the data consistency layer to enforce k-space data consistency [48]–[52]. It enables end-to-end methods to enjoy the benefits of enforcing data fidelity. Therefore, combining end-to-end and unrolled features in a single model may increase the diversity of network designs that also share the benefits of both categories.

## D.    Unsupervised Learning

The above discussion on unrolled vs end-to-end models assumes that the ground truth MR images are available to train the model to learn the mapping between the undersampled images and the ground truth. That is, the training process is supervised. If the ground truth images are not available, the model requires unsupervised training [53]. The objective is to minimise the difference between reconstructed images and the undersampled images at the undersampled k-space locations, i.e., enforcing data consistency [53]. Even without fully sampled ground truth, unsupervised models can remove undersampling artefacts effectively. The reason is that even without training, a deep learning model can capture a great deal of image statistics [54]. Most unsupervised methods use unrolled optimisation and alternately optimise the reconstructed images and the model parameters [24], [25], [55], [56]. Only one study implements an end-to-end training [53]. Most studies demonstrate higher quality reconstruction over traditional CS techniques [24], [25], [55], [56] and the supervised learning model [21] ADMM-CSNet [24]. This underscores the prospect of unsupervised learning when ground truth images are unavailable.

## II. NETWORK ARCHITECTURES

Having surveyed the two main categories of deep learning-based CS-MRI techniques, we will visit the key milestones during the development of deep learning-based CS-MRI (Supplementary Figure 3) (For less commonly used designs, refer to Supplementary Section I).

### A. Variational Network

Variational network (VN), an unrolled optimisation method, uses field-of-experts function as a regulariser in the image reconstruction loss function [30], [43]–[45] (2). Field-of-experts apply convolutional filters on the input undersampled images followed by activation functions. Unlike the activation functions used in typical neural networks, these functions are trainable. The trainable convolutional filters and activation functions are optimised to perform image regularisation (Supplementary Table 2). The strength of VN is that they require 10 to 100 times fewer parameters than a typical deep learning-based CS-MRI model. Therefore, the computational load may be lower with a smaller risk of overfitting. This has a greater potential for the more computationally demanding 3D or 4D reconstruction [45].

### B. Generative Adversarial Network

Generative adversarial network (GAN) has revolutionised the field of synthesising photo-realistic images [51], [57]. A GAN consists of a generator and a discriminator. The discriminator is trained to label the ground truth MR images as being 'real' and the reconstructed MR images as 'fake'. The generator does the opposite: its reconstructed images resemble fully sampled ones such that the discriminator would label the reconstructed images as 'real'. With an optimal discriminator, the generator minimises the Shannon-Jensen divergence between the reconstructed and fully sampled images (Supplementary Section II-E). De-aliasing GAN (DAGAN) [3] pioneers GAN-based CS-MRI, which consists of 10.8% of the models in this review [49], [51], [58]–[64]. DAGAN has achieved superior reconstruction performance compared with traditional CS techniques and ADMM-Net [65].

However, GAN suffers from training instability, slow convergence to the global minimum [66], [67], and vanishing gradient [68] . Wasserstein GAN (WGAN) could mitigate these issues [69]. Instead of the Shannon-Jensen divergence, WGAN minimises the Wasserstein distance between the reconstructed and fully sampled images. WGAN-based models outperform DAGAN and cycleGAN [60], [62], which do not use WGAN. Another limitation of GAN is that they 'over-emphasise the high-frequency texture and thus ignore image contents [51] and can produce over-smoothened appearance [70]. Least Square GAN (LSGAN) can address this problem [49], [63]. Taken together, while WGAN addresses the training instability of GAN, LSGAN may tackle the high-frequency texture issue.

Our interpretation of the effectiveness of GAN-based methods may be confounded by other model design elements. For example, in all GAN-based methods, the loss functions also penalise the deviations between reconstructed and fully sampled images in the image and/or k-space domain. Some studies penalise the perceptual quality difference using the VGG16 network [3], [60], others enforce k-space data fidelity [58]. Nevertheless, without other penalty terms, a GAN-only model still outperforms ADMM-Net [3]. Therefore, GAN-based techniques are promising CS-MRI reconstruction methods, whose performance can be further enhanced by auxiliary penalties.

## C. Input Domain

Among the studies in this review, 89.1% of the proposed models reconstruct the undersampled input in the image domain. Three [71]–[73] operate on the undersampled k-space, with higher reconstruction accuracy compared with the image domain techniques, e.g., DC-CNN and VN [72]. Two studies [32], [74] use a hybrid of k-space and image space. That is, for a 2D undersampled k-space input, inverse Fourier transform was performed along the x-axis. This means that the x-axis represents image domain information and y-axis k-space signals. The performance is higher over the image domain method ADMM-Net [74].

Cross-domain design, an increasingly popular strategy, leverages signals from multiple domains. KIKI-net pioneers cross-domain networks [8]. It concatenates a subnetwork operating on the k-space (k-net) with another subnetwork on the image domain (i-net), and so on. The undersampled k-space signals are first reconstructed by the k-net, followed by inverse-Fourier transform to the image domain to be processed by the i-net (Supplementary Figure 2). Similar network design follows [64], [75]–[77]. Apart from k-net and i-net, one study [40] concatenates a w-net, a subnetwork that operates on the wavelet-domain of the input image. The advantage of cross-domain network is that k-space-based network excels in removing high-frequency artefacts. Image-space network improves image sharpness and clarity [8]. Wavelet-domain network exploits both spatial and frequency features that may potentially accelerate feature learning [78]. Consequently, cross-domain networks outperform networks that operate only in the image domain [40], [64], [77], [79].

Besides joining subnetworks on different domains in series, some cross-domain networks concatenate subnetworks in parallel (Supplementary Figure 2) [38], [80]. The undersampled k-space signals are supplied to a k-net. In parallel, the undersampled image from the inverse Fourier transform is supplied to an i-net. Extensive connections between k-net and i-net facilitate the learning of a noise-free latent representation of the input image. This design surpasses image domain methods in reconstruction quality, including ADMM-CSNet (both studies), DC-CNN [80] and DAGAN [38]. However, cross-domain models are limited by their increased parameter numbers. To satisfy hardware requirement, each subnetwork in the IKIK-net needs to be trained separately [8]. This imposes challenges for training cross-domain models despite their exciting prospect by correcting undersampling artefacts from multiple domains [40].

## D.   Residual Learning

Residual learning (in 51.1% of deep learning-based CS-MRI designs) learns the difference or the residual between the ground truth and undersampled input, outperforming non-residual learning [3], [9]. The rationale is to 'constrain the generator to reconstruct only the missing details and prevent it from generating arbitrary features that may not be present in real MR images [3]. Residual learning can also mitigate training difficulty as the topological complexity of the residual difference may be smaller compared to the entire MR image [48]. This effectiveness has been justified by persistent homology analysis [17].

## E.   Attention

An attention module is a computational layer in the neural network [52], [82]. This module learns the most important pixel in the input to attend to, i.e., learning the optimal weights assigned to each pixel. Compared with the same model without attention modules, this design achieves a higher reconstruction accuracy. However, a key limitation of attention modules is their high computational demand, which is addressed by the memory-efficient self-attention module proposed by Zhou et al. [83].

# III. IMAGE REDUNDANCY

Another important technique to accelerate AI-powered fast MRI is image redundancy. Between 2019~2020, there has been a trend towards exploiting MR image redundancy across multiple contrasts, spatio-temporal dimensions, and parallel imaging coils, to improve performance and acceleration rates.

## A.   Redundancy Across Contrast Modalities

Different clinical settings demand different contrasts for MR images e.g., T1 weighted, T2 weighted, proton density etc. For example, T1 weighted images provide detailed anatomical structures while pathological features are usually more apparent in T2 weighted images [4]. To improve the clinical diagnostic power, MR images of multiple contrasts are required [84]. Because images with different contrasts of the same structure convey similar anatomical information, the information redundancy can be used to accelerate CS-MRI.

Among the reviewed studies, the earliest [85] uses fully sampled T1 weighted images to guide the reconstruction of the corresponding undersampled T2 weighted images. T1 weighted and T2 weighted images are concatenated as a 2-channel input to the deep learning model. This method achieves superior reconstruction performance compared with the model without the fully sampled T1 weighted image. A similar design was employed in [49]. Other studies [44], [86] concatenate undersampled images without the guidance of fully sampled ones. Alternatively, 2 separate networks are trained for separate contrasts with extensive cross-talk between the 2 networks, outperforming the same network without multi-contrast information [87]. However, the limitation of multi-contrast reconstruction is that signals from one contrast may leak into another [44]. Furthermore, the network cannot process an arbitrary number of contrasts without significant structural modifications. Despite these shortcomings, multi-contrast MR reconstruction represents a significant step forward in exploiting MR image redundancy.

## B.  Spatio-Temporal Redundancy

Spatio-temporal redundancy increases in higher-dimensional MR images. To illustrate, in 3D imaging, structures in 2 neighbouring planes are unlikely to be drastically different and are correlated. Likewise, in 4D imaging (3D spatial plus a temporal dimension), the structures between 2 adjacent time frames are correlated. However, extending 2D deep learning-based CS-MRI models to 3D and 4D usually requires computationally costly 3D or 4D convolution operations [88]. To mitigate the 3D computational demand, most studies [46], [79], [89]–[91] use 2+1 convolution. This involves a 2D convolution along two dimensions of the input image followed by 1D convolution along the rest of the one axis [92]. However, it is difficult to evaluate the performance of 3D deep learning-based CS-MRI models against typical deep learning methods, most of which (88.0% of the reviewed studies) target 2D reconstructions. Despite the computational challenge and the lack of evaluation frameworks, two studies [45], [93] venture into 4D MRI reconstruction. Analogous to 2+1 convolution, 3D spatial convolution followed by 1D temporal convolution is applied [93]. Hence, multi-dimensional MR image reconstruction tends to avoid computationally costly multi-dimensional convolutions.

## C.  Parallel Imaging with Coil Redundancy

In 41.3% of the reviewed studies, parallel imaging is combined with CS to exploit the k-space signal redundancy collected by multiple receiver coils. Similar to multi-contrast reconstruction, for separate imaging coils, many studies use separate input and outputs channels [17], [40], [58], [71], [76], [77], [79], [81], [94]–[96]. The reconstructed images for each coil are then combined by the sum-of-squares. One exception [95] uses a separate network to perform the coil combination. However, neither design can handle signals of an arbitrary number of coils.

Another approach is to incorporate parallel imaging into the optimisation objective. To illustrate, the coil sensitivity matrix $S_i$ describes the regions that a particular coil $i$ is most sensitive to. Then the image acquisition model (1) and the training objective (2) can be modified respectively as:

$$y_i = US_iFx,$$

and

$$\frac{1}{2}\big|\big|US_iFx - y\big|\big|_2^2 + R(x).$$

Deep learning models can be modified accordingly [12], [20], [26], [30], [43]–[45], [79], [93], [94], [97]–[102].

While deep parallel imaging CS techniques can further accelerate MR acquisition, evaluating their performance against single-coil reconstructions is challenging. This is because different datasets are required for multi- and single-coil applications. Alternately, coil compression of raw undersampled multi-coil data into single coil may be used but this comparison may not be fair [103]. Furthermore, in various multi-coil studies, coil compression of the multi-coil raw data into a smaller number of virtual coils was applied [41], [45], [91] to reduce the computational demands. It is unclear whether this measure can best utilise the multi-coil information or reflect the model performance on raw uncompressed multi-coil signals. Despite various computational and evaluational challenges to exploiting multi-contrast, spatio-temporal and parallel imaging redundancies, the recent developments reflect the remarkable community efforts in improving the speed and accuracy of CS-MRI.

# IV. META-ANALYSIS METHOD

## A. Data collection

To quantitatively evaluate the trend of deep learning-based CS-MRI development, we mined the literature across four platforms: google scholar, PubMed, IEEE and Crossref. We used the keyword: "MRI", "reconstruction" and "deep learning". The key word 'compressed sensing' was not incorporated to broaden the search range. The search was carried out on 22nd October 2020. The Publish or Perish software was used to search through google scholar, PubMed and Crossref. Xplorer was used for IEEE journals. The references of the matched studies were exported as a .ris file and imported into Mendeley Desktop. Using the 'Update Details' function in Mendeley Desktop, the details of all the references were updated automatically. To facilitate subsequent filtering statistical analysis, the references were transferred to Zotero, which enables the export the references as a .csv file. For consistency, Zotero was the reference manager for this paper.

### 1) Initial filtering

Initially, 1460 studies were identified that matched the three search keywords. Then 301 duplicates were removed based on a case insensitive match of the titles of the papers, leading to 1159 non-duplicated studies. Then the studies without titles or authors were removed, leaving 1144 studies for subsequent analysis. We excluded preprints, conference papers and other items that are not published in research journals for this review. This was done via a case insensitive search for the following keywords in the journal names and publishers of the papers:

"arxiv", "spie", "mirasmart", "proceeding", "patent", "openreview", "aaai", "conference", "book", "preprint", "meeting", "symposium", "workshop", "ismrm", "Proc Intl Soc Mag Recon Med", "icassp", "nips", "lectures", "book-chapter", "proceedings-article", "posted-content", "monograph", "dissertation"

After filtering, 578 studies remain for a title and abstract screening.

2)     Title Screening

Two independent reviewers determined the relevance of a research paper by screening its title. Our title screening criteria is that a study was removed if its title contains fewer than 2 of these three keywords: "compressed sensing", "MRI" and "deep learning". Any discrepancy between the two reviewers was resolved by the opinion of the more senior reviewer. After title screening, 221 studies entered the abstract screening stage.

3)     Abstract Screening

One reviewer performed abstract screening. The criterion was whether the abstract mentions all three of the keywords: "compressed sensing", "MRI" and "deep learning". If an abstract was generic, the introduction of the paper was briefly examined. Only 123 studies passed our abstract screening criteria.

4)     Full-text screening

Only studies with full text available in English and that proposed a new deep learning model for CS-MRI were included. The full text was screened by one reviewer and the results were scrutinised by a more senior reviewer. In the final review process, 92 studies were included. The entire process of literature screening and exclusion is summarised in Fig. 1a.

5)     Data collection

To summarise the key model design traits, criteria in the Checklist for Artificial Intelligence in Medical Imaging [104] were modified and tailored for deep CS-MRI studies (Table 2). For example, the modified checklist incorporated items that are salient for CS experiments but not necessarily for general -purpose imaging analysis. This includes the pattern of undersampling mask, acceleration ratio tested in the study, types of performance metrics etc. All data were collected by one reviewer and verified by another reviewer. Quantitative performance measures data were collected from the tables in the main text and supplementary files. Initial data cleaning was performed by the text editor vim and later using the programming language R.

**Table 2** Modified CLAIM criteria for collecting information from each paper during the meta-analysis.

| Category | CLAIM number[a] | Item | Explanation |
|---|---|---|---|
| Data | 7, 10 | Dataset | What the dataset(s) was, how it was collected, selection of subsets if appropriate |
| | | Region | Which region(s) of the body the images in the dataset covered |
| | | MRI sequence | What the MRI sequence was, e.g., T1, T2 etc |
| | 14 | Ground truth | What the ground truth was and how it was generated |
| | 20 | Partition | How the dataset was partitioned into training, validation, and testing subsets in terms of number of images, patients or MR scans |
| | 25 | Augmentation | How the training dataset was augmented |
| | 34 | Clinical feature | Whether the dataset contains images of pathology |
| Model | 22 | Category | What category the model belonged to, i.e., unrolled optimisation, end-to-end, or reference-driven |
| | 22 | Architecture | What the structure of the model was, e.g., U-Net, DC-CNN like etc |
| | | Channel number | How many coils the inputs signals to the model were |
| | | Channel merging | If the method uses multicoil input, i.e., a parallel imaging method, how the multicoil data was merged to produce a single final reconstructed image |
| | 22 | Input domain | Whether the model was designed to process raw k-space data, magnitude or complex image space data |
| | 22 | Dimension | What image dimension the model was designed to process, e.g., 2D spatial, 3D spatial, or 2D spatial-temporal |
| | | Input size | What size of the input MR images was, e.g., 256 x 256 |

| | 22 | Loss | Which loss function(s) were used to train the model |
|---|---|---|---|
| | 22 | Optimiser | Which optimiser was used to update and optimise network parameters, e.g., Adam, RMSProp |
| | 23 | Open source | Whether the link to the source code is mentioned in the main paper or supplementary file |
| | 23 | Platform | Which deep learning library was used to build the model, e.g., tensorflow, pytorch |
| Evaluation method | | Mask | What pattern(s) of undersampling mask was used, e.g., radial, variable density |
| | | Acceleration | Under which acceleration ratio(s) the model reconstructed the undersampled images |
| | | Comparison | What other algorithms were used to compare the performance of the proposed algorithm |
| | | Metric | What quantitative and qualitative metrics were used to evaluate reconstruction accuracy, e.g., NMSE, PSNR and SSIM |
| | 5 | Testing mode | Whether the model was tested prospectively (acquire the undersampled image via compressed sensing), retrospectively (undersampled after standard full acquisition or an established compressed sensing approach) or both |
| Result | 36 | Result | What were the quantitative data of performance metrics and/or qualitative comparison of representative reconstructed images |
| | | Computation time | Computation time in seconds on a GPU per reconstructed image |
| Discussion | | Novelty | What aspects of the model were novel, i.e., not previously reported |
| | | Strength | What problem in particular the model was designed to address |
| | 38 | Limitation | What problem remained in the model |

a The CLAIM item numbers that our proposed review criteria correspond to.

## B. Data analysis

### 1) Developmental trend

To determine changes in the popularity of deep CS-MRI model design traits, Pearson correlation was computed between the proportion of models using a particular trait and the year of publication. To assess whether numerical variables such as training or testing sample sizes change over time, the p-value was calculated. This was achieved by using the Kruskal-Wallis test implemented in the kruskal.test function from R package stats [105].

To evaluate the input image size, the pixel number of the input images to the models was calculated as followed. If more than 1 size were reported, the size of the image that was mentioned first in the paper was chosen. Then the width and height of the image were multiplied to obtain the pixel number. For studies that process 3D or 4D images, the dimension along the x and y axes are chosen instead.

To evaluate the reproducibility of different studies, the scoring was based on a previous review paper [106]. Studies that publish neither the code nor the dataset are classified as "hard to reproduce". Those that publish only the code or only the dataset are "medium to reproduce". Those that release both are "easy to reproduce".

### 2) Clustering

The following features were used to perform clustering analysis: GAN based, U-Net like, MSE loss, supervised, residual, complex input, parallel imaging, maximum acceleration, dimension, data consistency and spatial domain. Features not reported by most of the studies were excluded. The clustering algorithm was a Gaussian Mixture Model (GMM), implemented by the Mclust function from the R package mclust [107]. The number of mixture models was chosen to be between 1 to 20. Using Bayesian Information criteria, the optimal number of models was determined to be 8. The cluster names were annotated based on the design traits within each cluster.

To visualise the clusters, the principal component analysis was performed using the prcomp function in the R package stats. The means and variances of each Gaussian model cluster along all the features are projected onto the first two principal components and visualised as ellipses. The centres of the ellipses are the projected means. The axes lengths are twice the square root of variances, representing a 95% confidence interval.

The extent to which metrics are linearly correlated was quantified using the $R^2$ value. This was calculated on the author reported performance of a model at different acceleration ratios over 2 chosen metrics. The $R^2$ value was obtained using the lm function in R package stats.

3)    Performance

To quantify the improvement of a model over zero-filling reconstruction, the model performance at a particular metric was divided by the zero-filling performance, to obtain the odds ratio. If the performance data at more than one acceleration was available, the mean of the odds ratio was calculated. The Deeks' test for publication bias [108] was then carried out by calculating the p-value of the regression line between the odds ratio and one over square root of the effective sample size. In this study, the effective sample size was estimated using the testing sample size. The p-value was obtained using the stat_cor function in the R package ggpubr [109].

All the meta-analysis were performed using R version 3.6.3 running on Ubuntu 18.04. Upon the acceptance of this paper, we will release the source code (https://github.com/Yutong441/deepCSMRI) for reproducible and sustainable future studies.

# V. META-ANALYSIS RESULTS

Among the 92 studies that meet our meta-analysis inclusion criteria (Fig. 1a-b), the publication number increases exponentially from 2017 to 2020 (Fig. 1c). Most were from China and the USA, accounting for 63.0% of all studies (31.5% from either country) followed by Korea (12.0%). The institute with the highest number of publications was Stanford University, followed by Korea Advanced Institute of Science and Technology and Xiamen University (Fig. 1d). The rising publications underscores the increasing importance of deep learning-based CS-MRI.

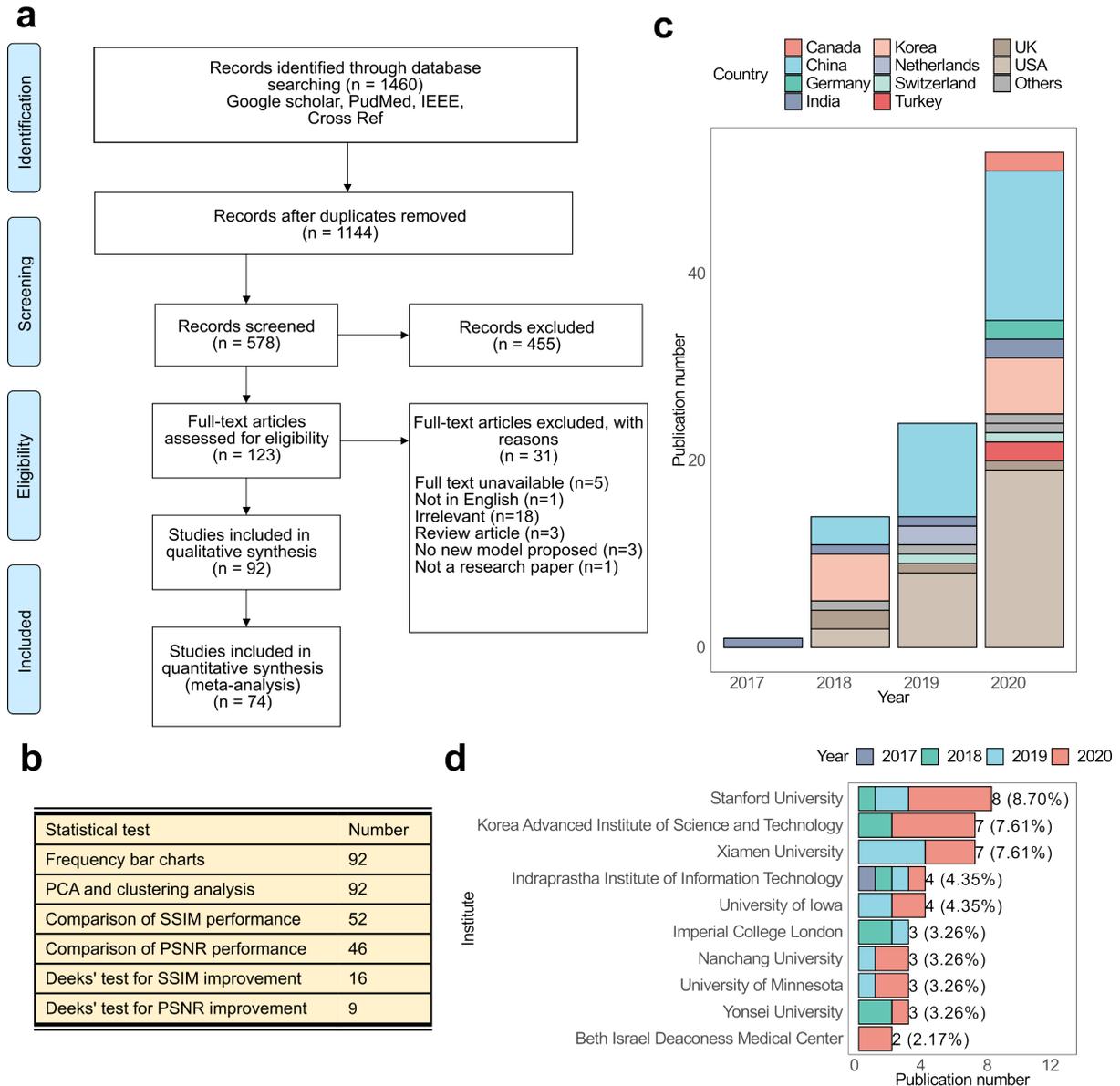

**Fig. 1. Basic information of the reviewed studies.**

**a**, Preferred Reporting Items for Systematic Reviews and Meta-Analyses (PRISIMA) [110] flow diagram of this meta-analysis study. **b**, The number of studies used in each analysis test. Abbreviations: PCA: principal component analysis; SSIM: structural similarity index, a popular metric of performance of deep learning model in CS reconstruction; PSNR: peak signal-to-noise ratio. **c**, The number of publications in each country over time. **d**, The number of publications by research institute. The number in the bracket indicates the proportion of the publications from the institute among all the studies that met the inclusion criteria (Section IV-A).

## A.    Dataset Characteristics

Training and evaluating deep learning-based CS-MRI models require the ground truth MR images. In 84.8% of the studies, fully sampled MR images served as the ground truth. However, in 15.2% of the studies, sampling MR images fully was impossible, for example, in cardiac cine imaging, due to the motion artefacts created by constant heartbeats [41]. Thus, CS-based reconstruction of these non-fully sampled images was treated as ground truth.

Regarding the type of datasets, forty-one studies (44.6%) used private datasets exclusively, while thirty (32.6%) exclusively used public datasets. Twenty-one (22.8%) used both public and private datasets. The most popular datasets were human connectome projects (used by 13.0% of the studies, Table 3), fastMRI (10.9%) and IXI (10.9%), but the tendency to use public datasets decreased over time (Fig. 2a).

Considering the sample size, the mean number of MR scans for model training was 89.9 and that for testing was 20.8. Neither the training nor testing sample size changed significantly over time (Fig. 2b-c). To increase the number of training samples, some studies applied data augmentation. The most popular augmentation techniques were flipping and rotation (Fig. 2d). Less popular techniques included adding random noise [88], sharpness, contrast [38], and using images of different acceleration ratios [111]. However, few studies assessed the impact of data augmentation on the performance of deep learning models in CS MRI.

 For the source of the datasets, most were collected from human volunteers or patients, except three from rats [24], [25], [48]. Most frequently, the spatial resolution of MR images were $256 \times 256$ MR (Fig. 2e). The lowest spatial resolution was $16 \times 16$ reflecting an image patch-based approach [112]. The highest spatial resolution was $590 \times 590$ [113]. Regarding the contrast of the MR images, T1 weighted and T2 weighted were the most popular (Fig. 2f). The least popular were magnetic resonance angiography (MRA), hyperpolarised $^{129}$Xe imaging (129Xe) and contrast-enhanced MRI (CE), probably linked to scarcity of publicly available datasets. Regarding the anatomical regions, the most popular were brain (45.45%) and knee (21.49%) (Fig. 2g), as most of the public datasets (Table 3) consisted of brain and/or knee images. Cardiac imaging was the third most popular (14.88%). Many (44.4 %) cardiac MRI-based studies utilised the temporal dimension of cardiac imaging, i.e., changes of cardiac MR images over each cardiac cycle. This included 66.9% of the studies that reconstruct 3D MR images and 100% of those using 4D images. Hence, the pursuit of multi-dimensional MRI reconstruction may fuel the interests towards cardiac MRI.

Concerning the pathological features of the datasets, 26.1% of the studies used pathology-free training and testing sets while 26.1% of the studies included pathology in both sets. Only 10 studies (10.9%) included pathology-free training sets and pathology-containing test sets to evaluate the generalisability of a deep learning model in pathology settings. Three studies demonstrated good generalisability [45], [102], [114].

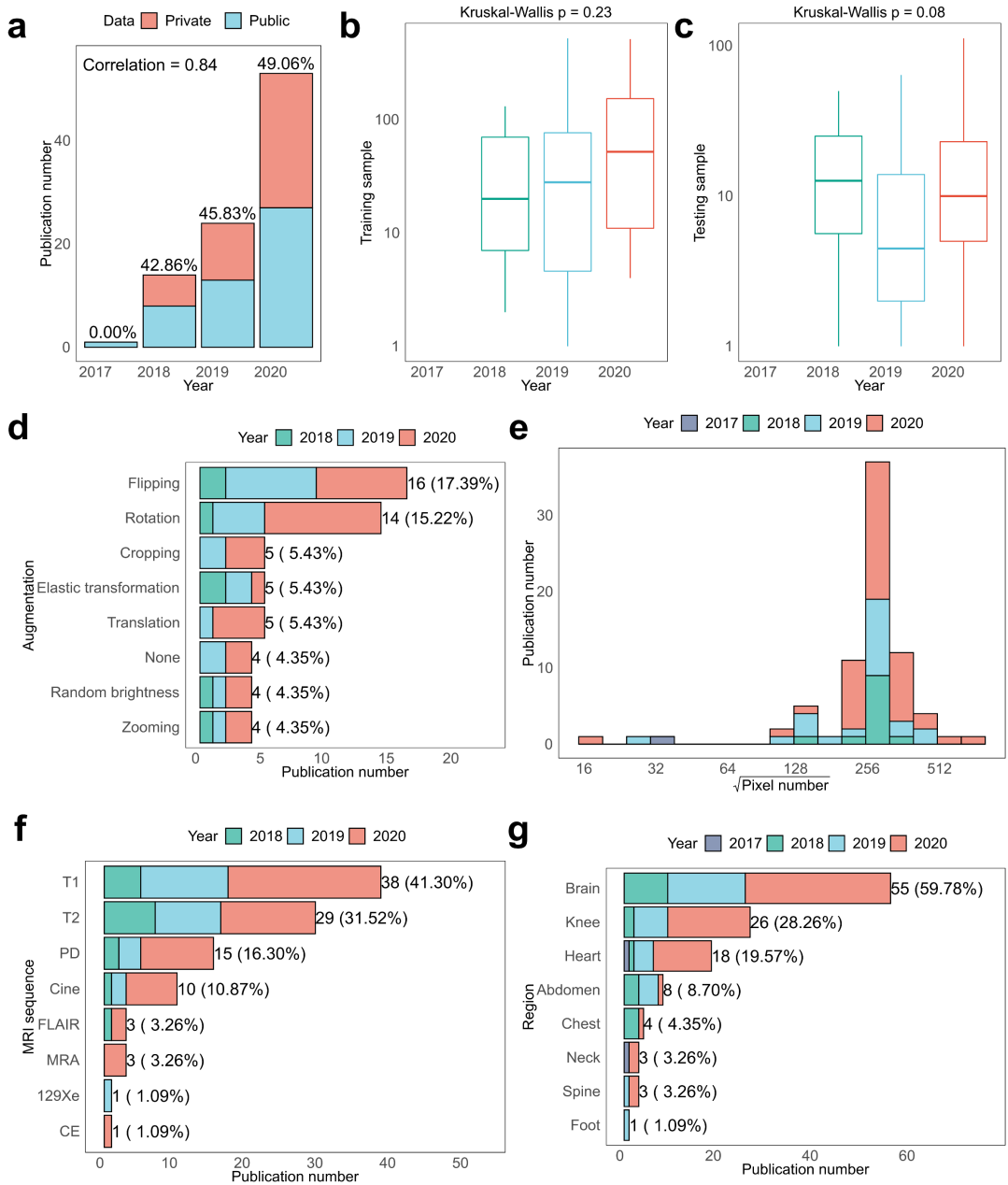

**Fig. 2. Features of the datasets used by different studies.**

**a**, The number of publications that use public and private datasets over time. The proportions of studies using public datasets are labelled above each bar. The correlation of these proportions with publication time was 0.839 (top left corner). **b-c**, Boxplot showing the training (B) and testing sample size (C) over time. The p-value of changes over time from Kruskal-Wallis test is shown above each plot. **d**, Histogram of the square root of pixel number of the input image to the deep learning models. For example, '256' indicates the input image is 256 × 256 in size. **e**, The MRI sequence of the images in the datasets used by different studies. Abbreviations: PD: proton density (including those with and without fat suppression); MRA: magnetic resonance angiography; 129Xe: 129Xe imaging; CE: contrast enhanced. **f**, The anatomical region of the images in the datasets used by different studies. **g**, The data augmentation techniques used by different studies.

**Table 3 Datasets used by each study.**

| Public dataset | Sample size | Pathology | Region | Used by |
|---|---|---|---|---|
| SRI24 [115] | 24 subjects[a] | No | Brain | [116] |
| MRBrainS13 [117] | 20 scans | Yes | Brain | [116] |
| ADNI [118] | NA[b] | Yes | Brain | [8, p. 201], [114] |
| NeoBrainS [119] | NA | NA | Brain | [116] |
| IXI | 600 images | No | Brain | [27], [35], [40], [49], [63], [64], [120]–[123] |
| fastMRI [124] | 8400 scans | NA | Brain, knee | [20], [47], [62], [81], [94], [100], [122], [123], [125] |
| Brainweb [126] | NA | Yes | Brain | [127] |
| mridata | 247 scans | NA | Knee | [72], [122], [127] |
| Caglary Campinas [128] | 212 scans | NA | Brain | [27], [28], [38], [76] |
| BRATS [129] | 300 scans | Yes | Brain | [48], [49], [121] |
| Dynamic MRI of speech movements | NA | NA | Brain, neck | [130] |
| MICCAI [131] | 47 scans | No | Brain | [3], [60], [132], [133] |
| HCP [134] | 9835 subjects | Yes | Brain | [12], [17], [28], [62], [72], [73], [81], [114], |

| | | | | [122], [135]–[137] |
|---|---|---|---|---|
| Aggarwal 2020 [47] | 10 subjects | NA | Brain | [47] |
| Aggarwal 2019 [22] | 5 subjects | NA | Brain | [22], [101] |
| OAI | NA | Yes | Knee | [138] |
| Hammernik 2018 [30] | 100 subjects | Yes | Knee | [26], [28], [30], [39], [47] |
| Vishnevskiy 2020 [45] | 18 subjects | Yes | Brain | [45] |
| MSChallenge [139] | 35 subjects | Yes | Brain | [64] |
| CAP [131] | 155 subjects | NA | Heart | [132] |
| Liu 2020 [140] | 31 scans | NA | Brain | [140] |
| Sawyer 2013 [141] | 20 scans | No | Knee | [48], [99] |
| MSSEG [142] | 53 scans | Yes | Brain | [85] |
| MRI MS DB | 100 subjects | Yes | Brain | [143] |
| MIDAS [144] | 58 subjects | No | Brain | [36], [49] |
| Liu 2019 [20] | 105 images | No | Brain | [20] |

[a] For the sample size, we report the size of the combined total datasets if more than one dataset are included on the website.

[b] Information not available from the website of the dataset.

## B.  Design

### 1)  Model Architecture

We also summarised the design traits of deep learning models. Supervised learning models and unrolled models were increasingly favoured, occupying greater proportions of the studies over time (Fig. 3a-b). Most studies (40.2%) use a U-Net like network while 7.6% used a structure similar to DC-CNN [10] (Fig. 3c). U-Net like networks became increasingly popular over time (correlation=0.87, Supplementary Table 3 and Section IV-B) while auto-encoder-based networks were less popular (correlation = -0.84). GAN-based models, data consistency layer and residual learning [10] were used in a considerable proportion of the studies. Both the data consistency layer and residual learning were increasingly incorporated (correlation=0.91 and 0.88 respectively).

### 2)  Loss Functions

Regarding the choice of loss functions (Fig. 3d), mean squared error (MSE) loss was the most frequently used, followed by L1 and L2 losses. Instead of MSE and L2, which can be over-smoothing [17], [95], some studies chose L1 loss [12], [49], [51], [73, p. 20] to facilitate convergence and produce sharper images [51]. To enforce data fidelity, some studies minimised data consistency loss, i.e., the difference between the undersampled k-space data and the reconstructed k-space at the undersampled locations. Some minimised the MSE in k-space, or in one study, in the wavelet domain [40]. To enhance perceptual quality, a few studies minimised the difference in the image embeddings from a trained VGG16 network between ground truth and reconstructed images [3]. Only two studies incorporated L2 regularisation as a strategy to prevent overfitting [36], [102, p. 20] and one use L1 regularisation [25]. One study [96] minimised the negative of SSIM of the reconstructed image, as SSIM was a key performance metric of CS reconstruction. Other performance metric-based loss functions included normalised MSE (NMSE) [133], normalised root MSE (NRMSE) [20] and mean absolute error (MAE) [35], [39]. For probabilistic deep learning models, the loss function was based upon maximum a posteriori [100] or the Kullback-Leibler divergence between the latent encodings for reconstructed and ground truth image distribution [59]. The loss functions with increased usage over time were L2, data consistency loss and RMSE (Supplementary Table 4). Altogether, MSE was the most prevalent loss function and recent deep learning-based CS-MRI developments have explored the diversity of loss function choices.

Various studies integrated multiple loss function to utilise the merits of them jointly (Fig. 3e). To illustrate, DAGAN [3] minimised the MSE in both image and k-space. The VGG16-based loss function was added to improve the perceptual quality. The GAN-based adversarial loss was integrated to generate photo-realistic images. Ablation experiments showed that each loss function was essential for DAGAN performance. Despite the advantages of multiple loss functions for the model training, its application did not change over time (correlation= 0.02). This may be because the training process was complicated by the need to balance the weightings of different loss components using the weighting hyperparameters.

Among the optimisers that apply the gradient of the loss function to update model parameters (Fig. 3c), the most used was the Adam (65.2%), with increasing popularity over time (correlation = 0.95), followed by the stochastic gradient descent (SGD) (9.8%). In contrast, RMSProp and gradient descent with momentum were used less frequently.

### 3) Input Characteristics

To process input MR image, the predominant method operated on 2D complex signals from the image domain (Fig. 3f). However, while raw MR signals are complex numbers, most deep learning frameworks do not support complex number calculations [103]. One solution is to only focus on the magnitude of the complex signals (14.1% of the studies). More commonly (60.9%), in the input layer of the neural network, one channel processed the real part of the MR signals, the other the imaginary part. Alternatively, the two channels can be used to process the magnitude and phase of the complex number signals [17], [49]. However, this magnitude-phase split has no benefits over the real-imaginary split [17]. Consequently, real-imaginary split dominates deep learning-based CS-MRI model designs.

However, such a real-imaginary split may not reflect the phase information of the complex signals [103]. To tackle this issue, complex convolution [39] convolves complex numbers using separate channels for real and imaginary images (Supplementary Section II-F), as adopted by subsequent studies [41], [75], [79], [93]. Complex convolution performance exceeded that of normal real-valued convolution [75] and networks that process magnitude images only [41]. However, calculations of complex numbers using real-valued channels may not be applicable for other computational layers of the neural networks, e.g., batch normalisation. The solution [75], [79] is radial batch normalisation, that is, performing batch normalisation on the magnitude image only, but the phase information is ignored. Despite attempts to circumvent complex-valued calculations, support for complex-valued operations is still an unmet need in deep learning frameworks.

### 4) Other Features

We next assessed the reproducibility of deep learning models by considering whether the dataset and source code were accessible [106]. 40.2% of the studies were 'hard to reproduce', which increased in proportion over time (correlation = 0.84, Fig. 3f). Moreover, computation time or parameter number did not change significantly over time (p=0.38, p = 0.50 respectively) (Fig. 3g-h).

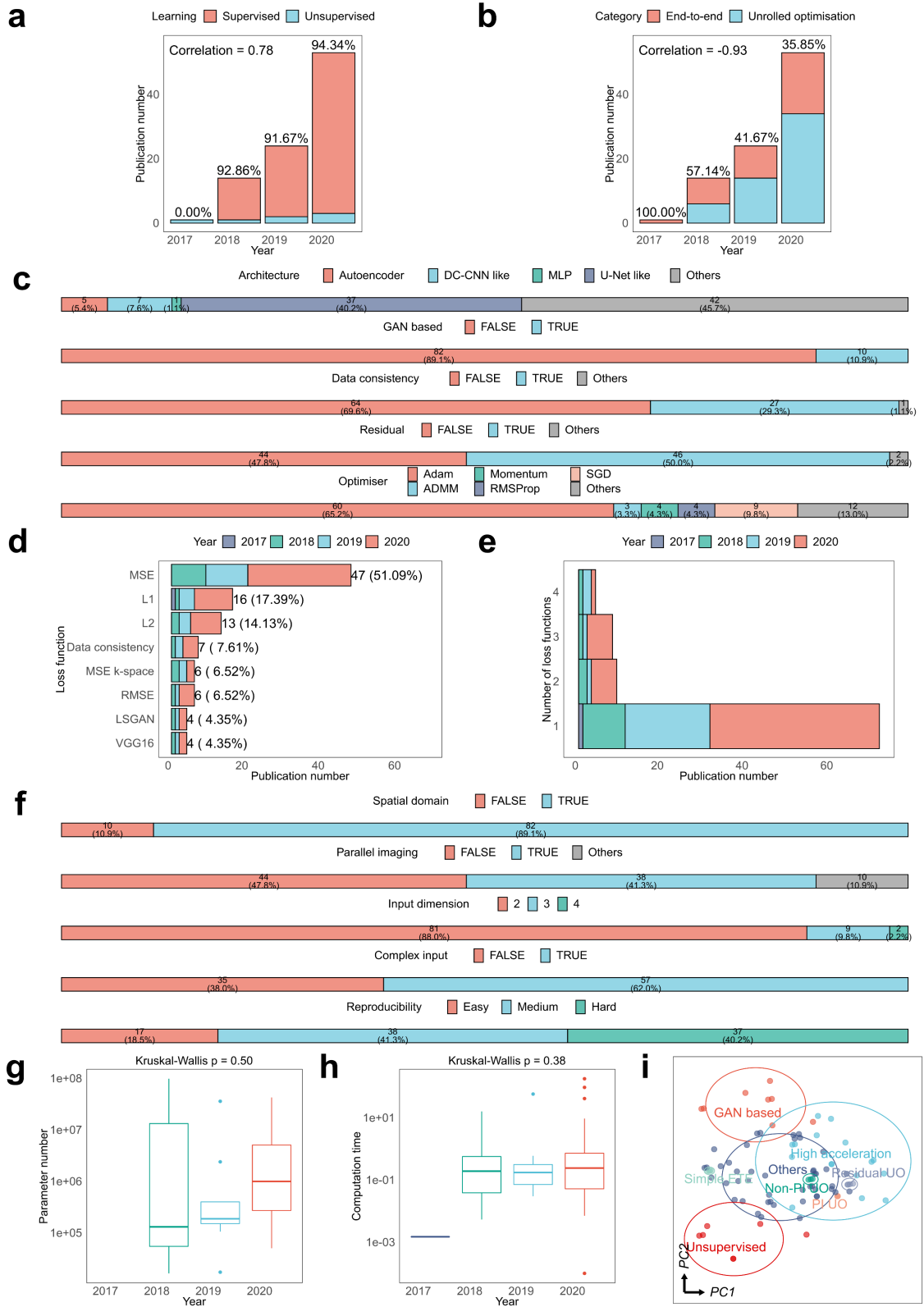

**Fig. 3. Neural network design traits.**

**a**, The number of publications over time by whether the model applies supervised learning. The proportion of unsupervised learning studies is in the text above each bar. **b**, The number of publications over time by model category. The proportion of end-to-end (ETE) models is indicated in the text above each bar. **c,** The number and proportion (in square brackets) of studies that use each of the key design traits in deep learning models. The field 'others' indicate the studies that did not report the relevant design traits. **d,** The loss functions used by different studies. Abbreviations: MSE: mean squared error between reconstructed image and ground truth; L1: L1 loss function; L2: L2 loss function; Data consistency: difference between reconstructed image and ground truth at undersampled k-space pixels; RMSE: root mean squared error; LSGAN: least-square generative adversarial network; VGG16: difference in the image embedding from VGG16 network between reconstructed image and ground truth. **e,** The number of loss functions used by different studies. **f,** Input image characteristics. **g,** Computation time over time. **h,** The number of parameters over time. The p value at the top was calculated using the Kruskal-Wallis test. **i,** Principal component analysis (PCA) of all the models. Each dot represents one model. The dots are coloured according to the Gaussian mixture model (GMM) cluster they belong to. The position and size of the ellipses represent the mean and twice standard deviation of each Gaussian model in the mixture, that are projected onto PC1 and PC2. Abbreviations: simple ETE: simple end-to-end models; PI UO: parallel imaging-based unrolled optimisation; Non-PI UO: unrolled optimisation without implementing parallel imaging; Residual UO: unrolled optimisation that implements residual learning.

Having characterised the model design traits, we used them to group the studies we reviewed into eight clusters. The GAN-based cluster featured the GAN-based CS-MRI models (Fig. 3i). The models within the 'high acceleration' cluster displayed high acceleration ratio and multi-dimensional MR image reconstruction. The 'unsupervised' cluster consisted of all unsupervised learning models. The 'simple ETE' cluster consisted of end-to-end (ETE) models that mostly do not implement data consistency layer, residual learning, GAN or parallel imaging. The unrolled optimisation (UO) models were subdivided into three clusters: residual UO that implemented residual learning, PI UO that implemented parallel imaging and non-PI UO. The 'others' cluster consisted of studies with diverse traits. We have therefore established an unbiased classification system to characterise the architectural traits of CS-MRI models.

## C.   Evaluation Metrics

We also compared how different studies evaluated the performance of their reconstruction models. CS-MRI models are tested by reconstructing undersampled images and comparing the reconstructed images with the ground truth. Undersampling can be retrospective, that is, undersampling the already acquired MR images. Prospective undersampling means collecting the undersampled k-space signals directly from the MR scanners and can better reflect performance in a real-life situation. Compared with prospective undersampling, retrospective undersampling is more financially and logistically feasible [5] and is implemented in 93.5% of the studies (Fig. 4a).

To compare the reconstructed images with the ground truth, most studies reported the structural similarity index measure (SSIM) and peak signal-to-noise ratio (PSNR) (Fig.

4b). Fewer used normalised root MSE (NRMSE), MSE, and signal-to-noise ratio (SNR). NRMSE and SSIM became more popular over time whereas PSNR and NMSE decreased in popularity (Supplementary Table 5). These metrics were quantitative, i.e., a defined algorithm that computes the similarity between the reconstructed and ground truth images. In contrast, qualitative metrics—measures without a clearly defined mathematical expression, including the rating scores by human observers and segmentation-based scoring—are used less frequently. The most popular qualitative metrics were image sharpness (6.52%), overall quality (4.35%), end-diastolic volume, ejection fraction and end-systolic volume as obtained by segmentation (3.26%) and Likert scale (3.26%). Thus, quantitative metrics such as SSIM, PSNR and NRMSE were the most prevalent.

Most of the studies reported at least two metrics to provide alternative performance quantifications (Fig. 4c). Across all the acceleration ratios and reported metric performance (Fig. 4d), we interrogated the redundancy of metrics by using one metric to predict the performance of another via linear regression. In Fig. 4e, the $R^2$ value between SSIM and PSNR is low, suggesting a non-linear relationship. For low model performance, i.e., low PSNR and SSIM values, SSIM was more sensitive to changes in model performance than PSNR. In contrast, for high performing models, PSNR was more sensitive. The results imply that PSNR and SSIM were unlikely a redundant pair of metrics, as each of them may be most sensitive to different performance levels.

Likewise, some quantitative metrics were not correlated, e.g., between SSIM, PSNR and MSE (Fig. 4f, Supplementary Table 7), though MSE, NRMSE, and NMSE are more closely related. The most highly correlated quantitative metric were are high frequency error norm (HFEN) and MAE, PSNR and MAE. Besides, many qualitative metrics were more linearly related, including sharpness, overall quality (OQ), artefact, contrast difference (CN) and contrast-to-noise ratio (CNR). Some quantitative and qualitative metrics also correlated, including RMSE and CNR, SNR and OQ, PSNR and OQ, and SNR and sharpness. However, this spuriously high similarity may arise because fewer studies reported qualitative than quantitative metrics. Despite the difficulty of interpreting the qualitative metrics, quantitative metrics e.g., SSIM, PSNR and MSE were not linearly dependent. This cautions future research against relying upon a single metric to assess model performance.

To fully evaluate the performance of a deep learning model, 51.9% studies compared the model performance with zero-filling, which represents the baseline reconstruction results. This involves filling the non-sampled k-space locations with zeros. Many studies also demonstrated the superiority of their models to other state-of-the-art techniques. Typical comparison techniques included U-Net like architectures, DC-CNN [10], ADMM-Net [65] and DAGAN [3] (Fig. 4g). Various studies also compared the performance to traditional techniques including DLMRI [42], TV [2], PANO [145], and BM3D [146]. Increasingly popular comparison methods included DC-CNN, U-Net and GRAPPA[147] but kt-SLR [148] was becoming less popular (Supplementary Table 6). Besides, 87.0% of the studies reported two or more comparison methods (Fig. 4h). Therefore, many studies benchmarked their models against zero-filling, U-Net, TV and DC-CNN.

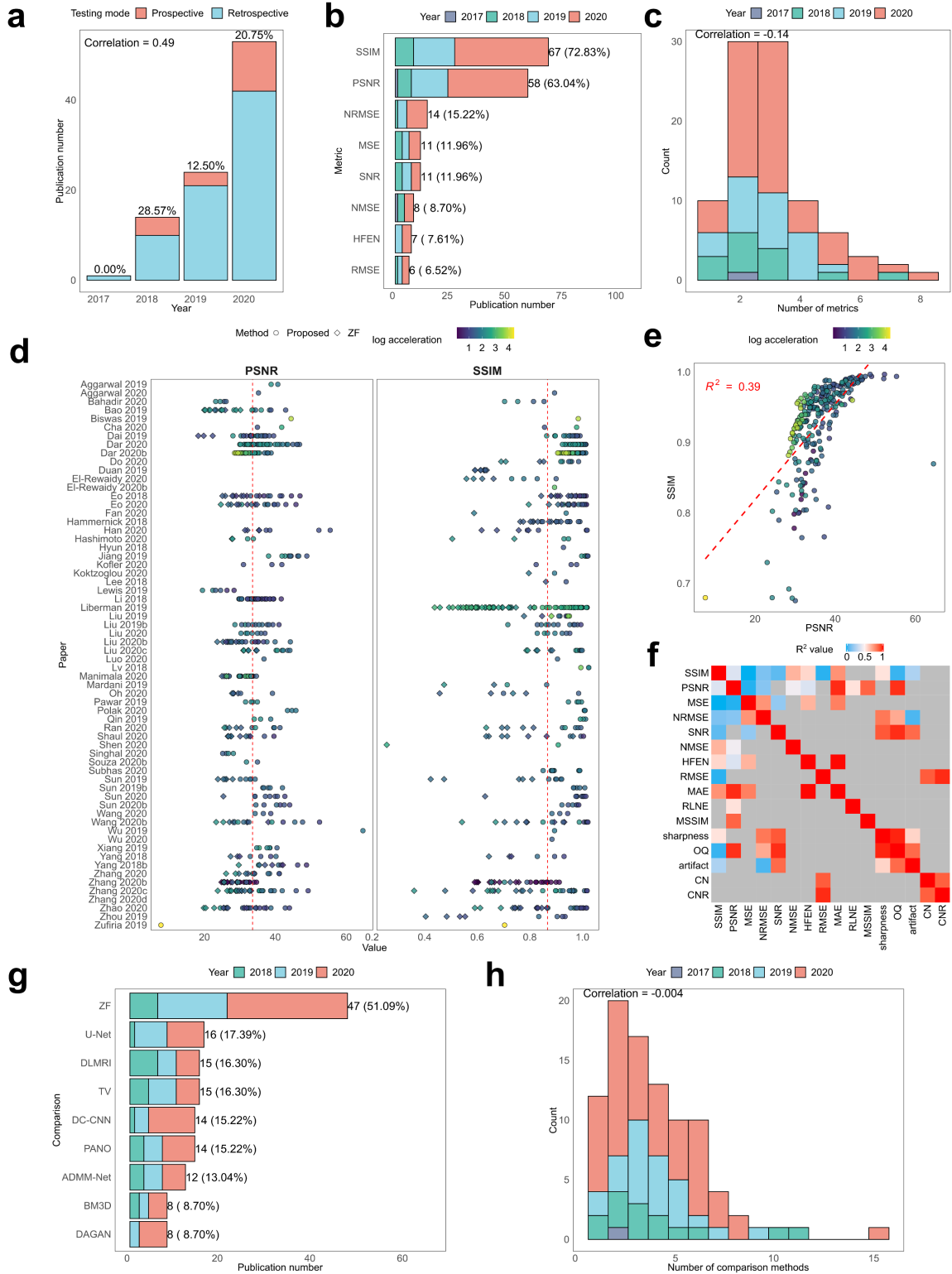

**Fig. 4. Evaluation of the model performance.**

**a**, The number of publications that applied prospective and retrospective undersampling over time. The proportion of models using prospective undersampling is in the text above each bar. **b**, The metrics used by different studies. Abbreviations: SSIM, structural similarity index; PSNR: peak signal-to-noise ratio; NRMSE: normalised root mean squared error; MSE: mean squared error; SNR: signal-to-noise ratio; NMSE: normalised mean squared error; HFEN: high frequency error norm; RMSE: root mean squared error. **c**, The number of metrics used by different studies to evaluate model performance. The correlation of the number of metrics with time is shown in the top left corner. **d**, PSNR and SSIM of the model performance across all the studies that provided performance statistics. Each dot represents the performance at a particular metric at a particular acceleration ratio using a particular method in a study. The red dashed line represents the mean of SSIM and PSNR across all studies over all acceleration ratios. **e**, The PSNR values of each model are plotted along the x axis and the corresponding SSIM values y axis. Each dot represents one model at a particular acceleration ratio. **f**, Heatmap showing the $R^2$ value between a pair of metrics. Gray region indicates where the $R^2$ value is not available due to small sample sizes. Abbreviations: MAE, mean absolute error; RLNE, relative L2 norm error; MSSIM, mean structural similarity index map; OQ: overall quality; CN: contrast difference; CNR: contrast-to-noise ratio. **g**, Comparison methods used by different studies. **h**, The number of comparison methods used by different studies. Abbreviations: ZF: zero filling; TV: total variation; DC-CNN: deep cascaded convolutional neural network; PANO: patch-based non-local operator; ADMM-Net: Alternating direction method of multipliers; DLMRI: dictionary learning MRI; BM3D: block matching 3D; DAGAN: de-aliasing generative adversarial network.

## D. Performance

Having assessed the model design traits and performance evaluation methods, we explored which design trait was associated with higher performance. We quantified the performance of each model by the odds ratio of improvement in either SSIM or PSNR over zero-filling. SSIM or PSNR odds ratios did not change significantly over time (Fig. 5a-b). On the reported improvements, the Deeks' test did not reveal publication bias (Fig. 5c-d). Across different clusters of models established earlier (Fig. 3i), the performance improvement was not significantly different (Fig. 5e-f). Furthermore, none of the design traits was significantly linked to performance improvement (Supplementary Table 8). The failure of detecting significant traits may be because performance comparison among different models was confounded by the disparity of dataset and evaluation metrics among them. Nevertheless, comparing the unadjusted p-values suggested that using Adam optimiser may lead to higher performance and using a U-Net like architecture and GAN may lead to a lower performance. A higher acceleration ratio was linked to higher SSIM improvement, probably because SSIM was the most sensitive to low performing models (Fig. 4e) and raising acceleration ratio tended to reduce performance.

Taken together, we have pioneered the meta-analysis framework, summarised the model design traits, analysed the developmental trend, and established a classification network for deep learning-based CS-MRI techniques, forming a comprehensive guide for future research.

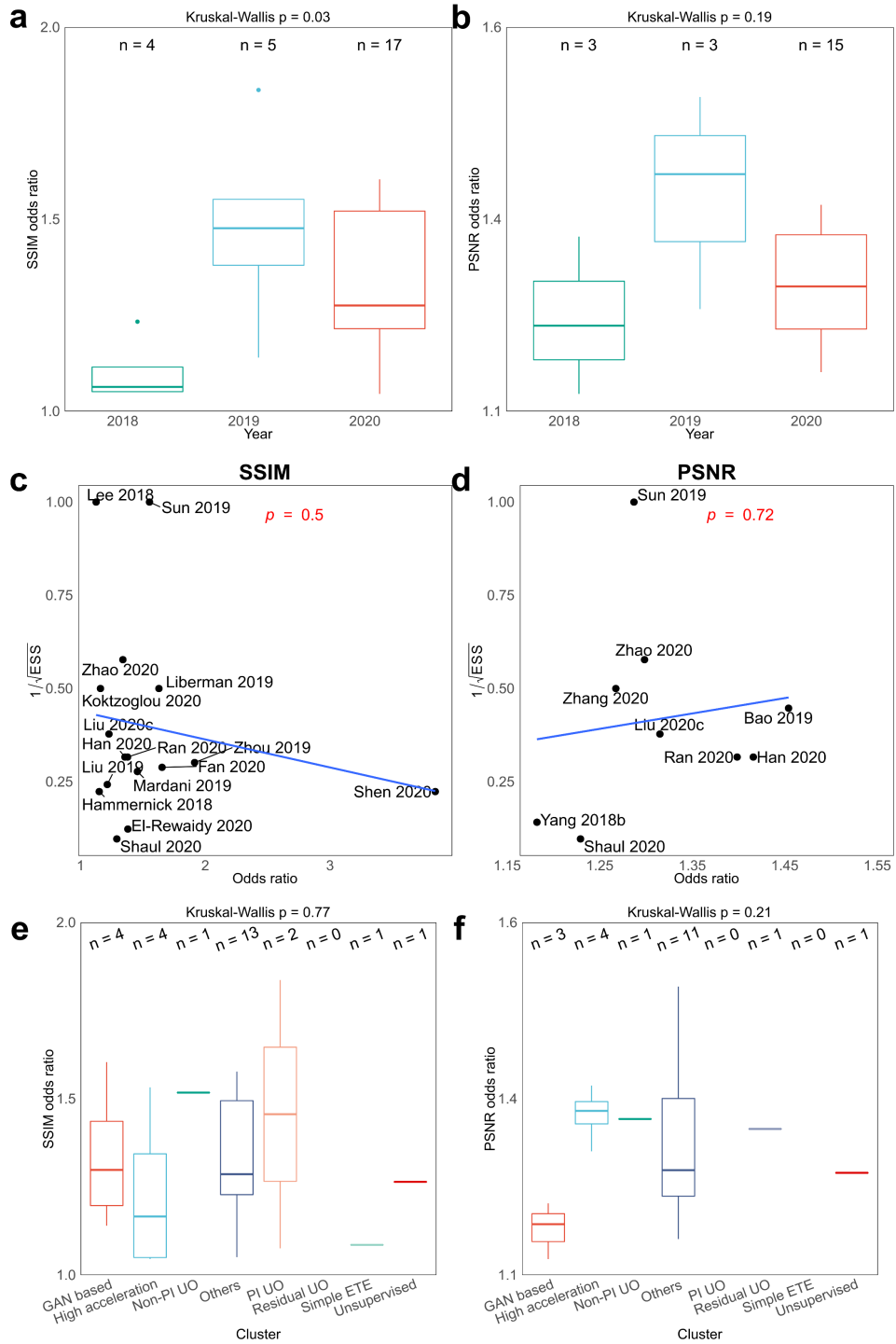

**Fig. 5. Performance improvement over the zero filled baseline reconstruction.**

**a**, The odds ratio of the SSIM improvement over time. The number of models assessed for each year group is shown above the boxes. **b**, The odds ratio of the PSNR improvement over time. **c**, Deeks' test using the SSIM odds ratio. **d**, Deeks' test using the PSNR odds ratio. **e**, The odds ratio of the SSIM improvement by clusters. **f**, The odds ratio of the PSNR improvement by clusters.

# VI.  CHALLENGES

Although deep learning-based CS-MRI techniques have advanced rapidly, they still suffer from limitations of the deep learning algorithms, most importantly, the dependency on large training data [11], [149]–[152]. Transfer learning may tackle this problem [36], [135]. With transfer learning, the models are trained on a source domain, in which training data are abundant, such as natural images. Subsequently, model parameters are fine-tuned in the target domain, in which the reconstruction is required but the training data is scarce. Using transfer learning, models trained with 4000 natural images perform as well as the same models trained with MR images [36]. Hence, transfer learning may address the demand for large training samples.

Another issue with deep learning-based CS-MRI models is their generalisability to different datasets or applications. A few studies report robust performance of the models across different datasets and noise levels [21], [27], [111], [133]. However, one study shows better performance in T1 weighted images compared with FLAIR MR images [8] and another displays higher reconstruction errors in fat-containing regions [153]. Furthermore, without transfer learning, deep learning models trained on natural or T1 weighted images cannot maintain equal performance on T2 weighted images. These results indicate that the same deep learning-based CS-MRI models may not display similar performance across different MR scanning sequences or anatomical regions. This is consistent with the instability of some deep learning-based methods e.g., VN, DAGAN and DC-CNN upon small perturbations of the input image [146].

While deep learning models have much shorter reconstruction time compared to traditional CS techniques [3], [10], they require a long training time [149], [155]. This is exacerbated by the need of hyperparameter tuning to select for the best performing models [43], [61], [71], [156], [157], as no theories currently govern deep learning model selection [149]. Despite their superior performance over traditional CS techniques, motion artefacts may not be effectively removed by deep learning models [9], [41], [125]. Therefore, computational challenges exist towards developing a universally applicable deep learning-based CS-MRI algorithm.

This systematic review and meta-analysis focused on how much a deep learning model improved beyond zero-filling reconstruction. However, such comparison is challenging given that different models were tested on different datasets, metrics, and acceleration ratios. Furthermore, we did not quantitatively explore the reconstruction time to identify the most computationally efficient models because not all models were implemented on the same computing platform. We also did not analyse the overfitting properties, i.e., the discrepancy in performance between the training and testing datasets. This is because most studies reported performance on testing, but not training datasets. Hence, to facilitate the future systematic review, we encourage future studies to test their model performance on commonly used datasets (human connectome projects, fastMRI, and IXI in Table 3) and metrics (PSNR, SSIM) and report performance on both training and testing datasets.

# VII. OUTLOOK

With the rapid rise of deep learning in computer vision, the past four years have witnessed substantial changes in the landscape of deep learning-based CS-MRI techniques. To summarise these developments, we have conducted systematic review and meta-analysis. We have introduced a comprehensive analysis framework based on the CLAIM criteria, summarised the typical deep neural network architectures in CS-MRI, compared their performance and evaluated their strengths and limitations. Earlier deep learning-based CS-MRI techniques have highlighted the developments of neural network architectures including data consistency layer, VN, GAN, residual learning, cross-domain networks etc. More recently, redundancy of MR images is explored, either across different contrasts, higher imaging dimensions or parallel imaging channels. However, with the increasing diversity of deep learning-based CS-MRI techniques, finding an appropriate and fair comparison benchmark is challenging. Nonetheless, we believe that the excitement of this field lies not only in improving beyond benchmark works but in creating new benchmarks for unexplored applications of CS-MRI. With this goal, efforts and drives among the deep learning community, milestones are set for faster and more accurate reconstruction performance. These developments may inspire other MRI applications such as MRI fingerprinting [158], [159], by synthesising a quantitative map of tissue properties from MR signal evolution over the signal acquisition trajectory. Therefore, we can envisage the development of deep learning-based fast CS-MRI will usher in a new era of digital healthcare and personalised medicine, which will be equipped with high throughput and low-cost imaging and more reliable quantitative imaging biomarker extraction and analysis.

# Supplementary Materials for "AI-based Reconstruction for Fast MRI—A Systematic Review and Meta-analysis"

Yutong Chen[1,2,3], Carola-Bibiane Schönlieb[4], Pietro Liò[5], Tim Leiner[6], Pier Luigi Dragotti[7], Ge Wang[8], Daniel Rueckert[9,10], David Firmin[1,2], and Guang Yang[1,2]

[1]National Heart & Lung Institute, Imperial College London, London SW3 6NP, U.K.
[2]Cardiovascular Research Centre, Royal Brompton Hospital, London SW3 6NP, U.K.
[3]Faculty of Biology, University of Cambridge, Cambridge CB2 1TN, U.K.
[4]Department of Applied Mathematics and Theoretical Physics, University of Cambridge, Cambridge CB2 1TN, U.K.
[5]Department of Computer Science and Technology, University of Cambridge, Cambridge CB2 1TN, U.K.
[6]Utrecht University Medical Centre, 3584 CX Utrecht, Netherlands
[7]Department of Electrical and Electronic Engineering, Imperial College London, London SW7 2BU, U.K.
[8]Biomedical Imaging Center, Rensselaer Polytechnic Institute, Troy, NY USA 12180
[9]Department of Computing, Imperial College London, London SW7 2BU, U.K.
[10]Institute for Medical Informatics, Statistics and Epidemiology, Technical University of Munich, 81675 Munich, Germany

## I. ADDITIONAL MODEL DESIGN TRAITS

This section supplements the main section 'Network Architecture' and details the less frequently used features in deep learning-based CS-MRI models.

### A. History Cognizance

In unrolled optimisation, the reconstructed image is updated iteratively. The reconstructed image from one iteration only depends on the input from the previous iteration. In a history cognizant model, the reconstructed image from one iteration is influenced by the results from all previous iterations. Recurrent Inference Machine (RIM) [1] and Convolutional Recurrent Neural Network (CRNN) [2] exemplify early history cognizant models. Both exploit the recurrent neural network (RNN) to achieve the history cognizant effect. In RNN, the input to a particular iteration of reconstruction is a weighted sum of inputs from previous iterations. The weights are learned to adjust the dependency on earlier and later iterations. To avoid the vanishing gradient problem of the RNN, both studies used a gated recurrent unit (GRU). While CRNN outperforms the non-history cognizant models such as DC-CNN, RIM demonstrates robustness to differing acceleration rates. A similar RNN based design [3] exceeds the performance of ADMM-CSNet, a non-history cognizant model. Apart from RNN, the attention module can achieve history cognizance [4]: The reconstructed images from all previous iterations are concatenated; The attention module learns which iteration contains the most salient feature for the final reconstruction. Hence, history cognizance can diversify the architecture and improve the performance of unrolled optimisation models.

### B. Learned Mask

Traditional CS and deep learning-based CS generally use the random undersampling mask of a predefined pattern, e.g., Gaussian, Poisson or radial, to satisfy the incoherency criteria of CS [5]. Two studies proposed learning the sampling distribution [6], [7]. The algorithm learns the optimal probability of sampling a particular pixel. This probability is constrained to between 0 and 1 using a logistic transformation. Based on the probability values, a Bernoulli sampling scheme determines whether to sample a particular pixel or not. Both studies reported improvements over deep CS-MRI methods without such undersampling optimisation.

### C. Uncertainty Quantification

Quantification of reconstruction uncertainty is valuable in discerning which structures from the undersampled images were reconstructed with high fidelity, and which with more uncertainty. It can assist clinical decision making of deep CS-MRI-accelerated images. To calculate the uncertainty values, one study [8] uses an auto-encoder based design. The authors introduced independent normal distributions to the latent encodings of the image. Then Monte Carlo sampling from this normal distribution forms the input to the decoder stream. Hence, from the same undersampled input, slightly different output reconstructions are generated. The mean across these outputs is the final image and the standard deviation quantifies uncertainty. Despite the value of uncertainty quantification, such a Monte Carlo sampling scheme raises the computational costs.

Instead of placing normal distribution on the latent representation, another technique used logistic distribution on the final reconstructed image [9]. However, the parameters for such logistic distributions are estimated using a deep neural network pixel-by-pixel iteratively, increasing the computational costs. A third approach leverages classification network [10]. Rather than reconstructing the images in floating numbers, floating numbers are discretised into 8-bit integers (256 integer values). The network classifies the probability of a reconstructed pixel belonging to each one of these 256 integer values. Hence, this network obtains a probabilistic distribution, i.e., uncertainty, for each pixel value. Despite the effectiveness of this method over ADMM-Net, discretising floating values reduces the numerical precision and increases the computational demand of the model [10]. Overall, uncertainty quantification in CS-MRI is computationally costly, which may deter widespread applications.

## II. MATHEMATICAL DERIVATIONS

### A. Bayesian View of Unrolled Optimisation

In the context of CS, the Bayes theorem describes the posterior probabilistic distribution of the reconstructed image x given the observed undersampled data y:

$$P(x|y) = \frac{P(y|x)P(x)}{P(y)},$$

where $P(x)$ is the prior distribution of the reconstructed image, $P(y)$ is the marginal likelihood of observing the undersampled data y, and $P(y|x)$ the conditional probability of measuring y if x is known. According to [9], [11], the objective for the reconstructed image is that the posterior probability of observing the reconstructed image x is maximised, i.e.,

$$x = \arg\max_x P(x|y).$$

Because $P(y)$ is independent of $x$,

$$x = \arg\max_x P(y|x)P(x).$$

This is the maximum a posteriori approach of CS reconstruction. Usually, the log posterior likelihood is maximised:

$$x = \arg\max_x \log P(y|x) + \log P(x).$$

The conditional probability $P(y|x)$ may be derived from the equation describing CS reconstruction:

$$y = Ax + \epsilon,$$

where $A = U\mathcal{F}$ represents the undersampled acquisition of the k-space data that correspond to the image x. The final undersampled acquisition is also assumed to be corrupted by some noise $\epsilon$. If the noise is Gaussian, i.e.

$$\epsilon \sim \mathcal{N}(0, \sigma^2).$$

Then $P(y|x)$ is also Gaussian:

$$P(y|x) \sim \mathcal{N}(Ax, \sigma^2),$$

and

$$\log P(y|x) = -\frac{||y - Ax||^2}{2\sigma^2} - \frac{1}{2}\log(2\pi\sigma^2).$$

Because $\sigma$ is assumed to be independent of x,

$$x = \arg\max_x -\frac{||y - Ax||^2}{2\sigma^2} + \log P(x),$$

which is equivalent to,

$$x = \arg\min_x \frac{||y - Ax||^2}{2\sigma^2} - \log P(x).$$

Using the notation convention of CS problem:

$$x = \arg\min_x \frac{\lambda}{2}||y - Ax||^2 - \log P(x),$$

this equation matches the general problem for CS in which the first term, derived from the conditional probability expression, enforces k-space data consistency. The second term, derived from the prior distribution of the reconstructed image x, matches the regulariser term of CS. Hence, the regulariser of CS represents prior domain-specific knowledge about the expected property of the reconstructed image.

*B. Solving the CS Problem with Gradient Descent*

Many unrolled optimisation models are implemented in an end-to-end fashion to facilitate the training process. This is achieved by expressing the general CS problem in a closed form which can be treated as a computational layer in the neural network:

$$\arg\min_x \mathcal{L} = \arg\min_x \frac{\lambda}{2}||y - Ax||_2^2 + R(x),$$

where $\mathcal{L}$ is the loss function or training objective. A few studies especially those related to variational network (VN), used gradient descent to express this loss function in a closed form and train the model in an end-to-end fashion. In each iteration of gradient descent, the reconstructed image is updated along the gradient of its training objective or loss function:

$$x_{t+1} = x_t - \alpha \nabla_{x_t} \mathcal{L}(x_t),$$

where $x_t$ is the reconstructed image at the $t^{th}$ iteration and $\alpha$ adjusts the size of gradient descent. In VN for example, the loss function is

$$\mathcal{L}(x_t) = \frac{\lambda}{2}||y - Ax||_2^2 + \sum_i f_i(k_i x_t).$$

Similar to the general CS problem, the first term of the loss function enforces data consistency. In the second term, the trainable convolutional filters $k_i$ and activation functions $f_i$ learn the optimal way to regularise the image towards reconstruction. Computing the gradient is trivial, leading to the final closed form expression:

$$x_{t+1} = x_t - \alpha(\lambda A^*(Ax - y) + \sum_i k_i^T f_i'(k_i x_t)),$$

where $f_i'$ is the first derivative of the activation function and $A^*$ the complex conjugate of A.

*C. Solving the CS Problem with Conjugate Gradient Descent*

In addition to gradient descent, conjugate gradient descent can be used to solve the general CS problem [6], [12], [13]. Here we illustrate the use of conjugate gradient descent using the algorithm from [13]. The training objective is:

$$\arg\min_x \mathcal{L} = \arg\min_x ||y - Ax||_2^2 + \frac{\lambda_1}{2}||x - f(x|\theta)||_2^2 + \lambda_2(\text{Tr}(x^T Dx) - 2\text{Tr}(x^T Wx)),$$

where the first term enforces data consistency, and the second term directs the final reconstructed image towards the output of a neural network $f(x|\theta)$. The third term is the SToRM priors that exploit non-local redundancies between images. Terms $\lambda_1$ and $\lambda_2$ are scalars that tune the relative contributions of these three terms to final image reconstruction. Using the alternating minimisation technique, at the $t^{th}$ iteration, the authors first keep the reconstructed image x constant and calculated the value of the following terms:

$$Y = f(x_t|\theta),$$

and

$$Q = Wx_t.$$

Then the conjugate gradient of the loss function with respect to $x$ was computed:

$$\nabla_x \mathcal{L} = A^*(Ax - y) + \lambda_1(x - Y) + \lambda_2(Dx - Q) = 0.$$

The optimal reconstruction occurs when the conjugate gradient of the loss function equals to 0. Solving for $x$ yields

$$x = (A^*A + \lambda_1 \mathbb{I} + \lambda_2 D)^{-1}(A^*y + \lambda_1 Y + \lambda_2 Q).$$

This equation hence expresses the loss function in a closed form and allows it to be integrated as a computational layer in the neural network.

*D. Solving the CS Problem with Auxillary Variables*

Apart from gradient and conjugate gradient descent, the CS problem can be solved by introducing auxiliary variables into the equation, in order to break down the general CS problem into multiple sub-problems, each of which is simpler to solve. This is used in IFR-Net [14], ADMM related algorithms [15], [16] and deep dictionary learning [17], [18]. In general, an auxiliary variable $u$ is introduced to the regulariser term as:

$$\min_{x,u} \frac{\lambda}{2}||y - Ax||_2^2 + R(u) \ s.t. \ x = u.$$

Then a quadratic penalty can be used to relax the equality constraint [19]:

$$\min_{x,u} \frac{\lambda}{2}||y - Ax||_2^2 + R(u) + \frac{\rho}{2}||x - u||_2^2,$$

in which the last term minimises the difference between the auxiliary variable and the reconstructed image x. This problem can then be solved by the alternating direction method of multipliers (ADMM), which breaks down the problem into 2 subproblems. In each subproblem, one variable is optimised while the other is kept constant:

$$u = \arg\min_{u} R(u) + \frac{\rho}{2}||x - u||_2^2,$$

$$x = \arg\min_{x} \lambda||y - Ax||_2^2 + \rho||x - u||_2^2.$$

However, the solutions to the above problems vary depending on the regulariser used in each deep learning application. The subproblem for auxiliary variable $u$ can be solved in closed form [15], using gradient descent [14] or conjugate gradient descent [20]. Some algorithms [15], [17], [18] introduce multiple auxiliary variables that can be updated in similar manners. The subproblem for x can be solved by the Tikhonov regularisation:

$$x = (A^*A + \rho\mathbb{I})^{-1}(A^*y + \rho u).$$

*E. Generative Adversarial Network*

In the generative adversarial network (GAN), the discriminator is trained to distinguish fully sampled ground truth $\hat{x}$ from reconstructed images x by classifying them as real and fake respectively. The generator is trained to 'fool' the discriminator so that it cannot tell the difference between $\hat{x}$ and $x$. The entire process is a minimax game described as:

$$\mathcal{L}(\theta_D, \theta_G) = \min_{\theta_G}\max_{\theta_D}\mathbb{E}_{\hat{x}\sim p_{data}(\hat{x})}[\log D_{\theta_D}(\hat{x})] + \mathbb{E}_{x\sim p(x)}[\log(1 - D_{\theta_D}(G_{\theta_G}(x)))],$$

in which $\theta_D$ and $\theta_G$ are the parameters in the discriminator and generator respectively, $p_{data}$ is the distribution of the fully sampled ground truth and $p(x)$ the distribution of the reconstructed images. The first term maximises the likelihood that the discriminator will classify the ground truth as real. The second term maximises the likelihood that the discriminator will classify the reconstructed image as fake, but also that the generator will create a real image. Assuming $G$ is fixed, the optimal discriminator [21] is:

$$\theta_D = \frac{p_{data}(\hat{x})}{p_{data}(\hat{x}) + p_G(x)}.$$

Therefore, the objective of GAN can be written as:

$$\mathcal{L}(\theta_\mathcal{G}) = 2\mathrm{JSD}(p_{data}||p_G) - 2\log 2.$$

Hence, GAN minimises the Jensen-Shannon divergence (JSD) between the distributions of the ground truth image and that of the reconstructed image. Because of the training instability of GAN, Wasserstein GAN was proposed with the following training objective:

$$\mathcal{L}(\theta_D, \theta_G) = \min_{\theta_G}\max_{\theta_D}\mathbb{E}_{\hat{x}\sim p_{data}(\hat{x})}[D_{\theta_D}(\hat{x})] - \mathbb{E}_{x\sim p(x)}[D_{\theta_D}(G_{\theta_G}(x))].$$

This loss function minimises the Wasserstein distance between the ground truth image and reconstructed image provided the discriminator satisfies the 1-Lipschitz constraint. For further information about why the above loss function minimises the Wasserstein distance and how the 1-Lipschitz constraint is enforced, we refer readers to the original WGAN paper [22].

*F. Complex Convolution*

To circumvent the problem of lack of complex number supports in certain deep learning frameworks, complex convolution was proposed [23], [24]. It performs a convolution of two complex tensors by decomposing the complex numbers into their real and imaginary components:

$$w * x = (w_{real} + iw_{imaginary}) * (x_{real} + ix_{imaginary}).$$

Expanding the brackets:

$$w * x = (w_{real} * x_{real} - w_{imaginary} * x_{imaginary}) + i(w_{real}x_{imaginary} + w_{imaginary}x_{real}).$$

## III. Supplementary Tables

**Supplementary Table 1** Comparison between end-to-end and unrolled optimization algorithms.

| Feature | End-to-end | Unrolled |
|---|---|---|
| Training sample | Large | Small |
| Weight updates | Yes | Yes |
| Image update | No | No |
| CS specific | No | Yes |
| Computation time | Short | Long |

**Supplementary Table 2** Comparison of the features of the variational network (VN) models. Abbreviations: RBF, radial basis function; MSE, mean squared error

| Study | Number of filters | Activation function | Number of activation functions | Loss function |
|---|---|---|---|---|
| Hammernick 2018[25] | 48 | RBF | 31 | MSE |
| Chen 2018[26] | 20 | RBF | 31 | MSE |
| Vishnevskiy 2020[27] | 8 | linear | 4 | L1 |
| Polak 2020[28] | 24 | RBF | 31 | L2 |

**Supplementary Table 3** Correlation of the use of neural network design traits over time.

| Feature | Variable | Correlation |
|---|---|---|
| Parallel imaging | TRUE | 0.98 |
| Optimiser | Adam | 0.95 |
| Unrolled | TRUE | 0.93 |
| Data consistency | TRUE | 0.91 |
| Residual | TRUE | 0.88 |
| Architecture | U-Net like | 0.87 |
| Complex input | TRUE | 0.85 |
| Reproducibility | Hard | 0.84 |
| Data | Public | -0.84 |
| Architecture | Autoencoder | -0.84 |
| Reproducibility | Medium | -0.81 |
| Architecture | Others | 0.80 |
| Learning | Supervised | 0.78 |
| Input dimension | 2 | -0.70 |
| Code | Public | 0.70 |
| Reproducibility | Easy | 0.69 |
| Spatial domain | TRUE | -0.55 |
| GAN based | TRUE | 0.54 |
| Optimiser | Momentum | 0.40 |
| Architecture | DC-CNN like | 0.27 |
| Optimiser | ADMM | 0.14 |
| Optimiser | RMSProp | -0.08 |

**Supplementary Table 4** Correlation of the use of different loss functions over time.

| Loss function | Correlation |
|---|---|
| L2 | 0.84 |
| Data consistency | 0.83 |
| RMSE | 0.78 |
| L1 | -0.72 |
| MSE | 0.68 |
| LSGAN | 0.43 |
| VGG16 | 0.43 |
| MSE k-space | 0.15 |

**Supplementary Table 5** Correlation of the use of different metrics over time.

| Metric | Correlation |
|--------|-------------|
| NRMSE | 0.92 |
| SSIM | 0.83 |
| NMSE | -0.83 |
| PSNR | -0.67 |
| RMSE | 0.55 |
| MSE | 0.21 |
| SNR | 0.19 |
| Sharpness | 0.09 |

**Supplementary Table 6** Correlation of the use of different comparison methods over time.

| Comparison | Correlation |
|-----------|-------------|
| DC-CNN [29] | 1.00 |
| Zero-filling | 0.81 |
| kt-SLR [30] | -0.76 |
| U-Net like | 0.69 |
| GRAPPA [31] | 0.63 |
| PANO [32] | 0.53 |
| BM3D [33] | 0.41 |
| ADMMNet [16] | 0.36 |
| TV [5] | 0.27 |
| FDLCP [34] | 0.14 |
| TL [35] | 0.14 |
| PICS [5] | -0.08 |
| SIDWT [36] | -0.08 |
| DLMRI [37] | 0.04 |
| PBDW [38] | 0.01 |

**Supplementary Table 7** $R^2$ value for pairwise metric comparison. NA: value not available as fewer than three studies report the metric pairs.

| Metric | SSIM | PSNR | MSE | NRMSE | SNR | NMSE | HFEN | RMSE | MAE | RLNE | MSSIM | Sharpness | OQ | Artefact | CN | CNR |
|--------|------|------|-----|-------|-----|------|------|------|-----|------|-------|-----------|-----|----------|-----|-----|
| SSIM | 1.00 | 0.39 | 0.00 | 0.14 | 0.08 | 0.67 | 0.58 | 0.02 | 0.76 | NA | NA | 0.54 | 0.03 | 0.26 | NA | NA |
| PSNR | 0.39 | 1.00 | 0.01 | 0.16 | NA | 0.46 | 0.39 | NA | 0.97 | 0.55 | 0.87 | NA | 0.97 | NA | NA | NA |
| MSE | 0.00 | 0.01 | 1.00 | 0.76 | 0.19 | NA | 0.67 | NA | 0.81 | NA | NA | NA | NA | NA | NA | NA |
| NRMSE | 0.14 | 0.16 | 0.76 | 1.00 | NA | NA | NA | NA | NA | NA | NA | 0.83 | 0.70 | 0.06 | NA | NA |
| SNR | 0.08 | NA | 0.19 | NA | 1.00 | NA | NA | NA | NA | NA | NA | 0.90 | 0.97 | 0.86 | NA | NA |
| NMSE | 0.67 | 0.46 | NA | NA | NA | 1.00 | NA | NA | NA | NA | NA | NA | NA | NA | NA | NA |
| HFEN | 0.58 | 0.39 | 0.67 | NA | NA | NA | 1.00 | NA | 1.00 | NA | NA | NA | NA | NA | NA | NA |
| RMSE | 0.02 | NA | NA | NA | NA | NA | NA | 1.00 | NA | NA | NA | NA | NA | NA | 0.89 | 0.97 |
| MAE | 0.76 | 0.97 | 0.81 | NA | NA | NA | 1.00 | NA | 1.00 | NA | NA | NA | NA | NA | NA | NA |
| RLNE | NA | 0.55 | NA | NA | NA | NA | NA | NA | NA | 1.00 | NA | NA | NA | NA | NA | NA |
| MSSIM | NA | 0.87 | NA | NA | NA | NA | NA | NA | NA | NA | 1.00 | NA | NA | NA | NA | NA |
| Sharpness | 0.54 | NA | NA | 0.83 | 0.90 | NA | NA | NA | NA | NA | NA | 1.00 | 0.98 | 0.58 | NA | NA |
| OQ | 0.03 | 0.97 | NA | 0.70 | 0.97 | NA | NA | NA | NA | NA | NA | 0.98 | 1.00 | 0.86 | NA | NA |
| Artefact | 0.26 | NA | NA | 0.06 | 0.86 | NA | NA | NA | NA | NA | NA | 0.58 | 0.86 | 1.00 | NA | NA |
| CN | NA | NA | NA | NA | NA | NA | NA | 0.89 | NA | NA | NA | NA | NA | NA | 1.00 | 0.88 |
| CNR | NA | NA | NA | NA | NA | NA | NA | 0.97 | NA | NA | NA | NA | NA | NA | 0.88 | 1.00 |

**Supplementary Table 8** Impact of model design traits on model performance. **Method**: statistical test to calculate the p value, Wilcoxon: Wilcoxon rank sum test, Pearson, Pearson correlation. **Absence**: the mean of the performance metric improvement relative to zero-filling reconstruction in the studies that do not implement the particular design. **Presence**: the mean of the performance metric improvement in the studies that implement the design. **Absence number**: the number of studies that do not implement the particular design. **Presence number**: the number of studies that implement the particular desing. **P-value**: unadjusted p-values. **Adjusted**: p-values adjusted by Benjamini-Hochberg method.

| Metric | Feature | Variable | Method | Absence | Presence | Absence number | Presence number | P-value | Adjusted |
|--------|---------|----------|--------|---------|----------|----------------|-----------------|---------|----------|
| SSIM | Optimiser | Adam | Wilcoxon | 1.22 | 1.58 | 9 | 17 | 0.01 | 0.23 |
| PSNR | Architecture | U-Net like | Wilcoxon | 1.35 | 1.25 | 13 | 8 | 0.01 | 0.29 |
| SSIM | Maximum acceleration | NA | Pearson | NA | NA | NA | NA | 0.02 | 0.45 |
| PSNR | GAN based | TRUE | Wilcoxon | 1.33 | 1.21 | 18 | 3 | 0.02 | 0.45 |
| PSNR | Cluster | GAN based | Wilcoxon | 1.33 | 1.21 | 18 | 3 | 0.02 | 0.45 |
| PSNR | Cluster | High acceleration | Wilcoxon | 1.30 | 1.36 | 17 | 4 | 0.12 | 1.00 |
| SSIM | Spatial domain | TRUE | Wilcoxon | 1.27 | 1.50 | 5 | 21 | 0.20 | 1.00 |
| SSIM | Cluster | High acceleration | Wilcoxon | 1.49 | 1.27 | 22 | 4 | 0.28 | 1.00 |
| SSIM | Loss function | MSE | Wilcoxon | 1.34 | 1.55 | 12 | 14 | 0.32 | 1.00 |
| PSNR | Spatial domain | TRUE | Wilcoxon | 1.33 | 1.31 | 5 | 16 | 0.40 | 1.00 |
| SSIM | Cluster | Others | Wilcoxon | 1.35 | 1.56 | 13 | 13 | 0.42 | 1.00 |
| PSNR | Data consistency | TRUE | Wilcoxon | 1.30 | 1.33 | 13 | 8 | 0.46 | 1.00 |
| PSNR | Residual | TRUE | Wilcoxon | 1.31 | 1.32 | 8 | 13 | 0.50 | 1.00 |
| PSNR | Optimiser | Adam | Wilcoxon | 1.32 | 1.31 | 7 | 14 | 0.58 | 1.00 |
| PSNR | Loss function | MSE | Wilcoxon | 1.30 | 1.32 | 8 | 13 | 0.70 | 1.00 |
| PSNR | Category | Unrolled optimisation | Wilcoxon | 1.33 | 1.30 | 7 | 14 | 0.74 | 1.00 |
| PSNR | Parallel imaging | TRUE | Wilcoxon | 1.32 | 1.30 | 16 | 4 | 0.75 | 1.00 |
| SSIM | Complex input | TRUE | Wilcoxon | 1.39 | 1.48 | 7 | 19 | 0.78 | 1.00 |
| SSIM | Category | Unrolled optimisation | Wilcoxon | 1.54 | 1.39 | 11 | 15 | 0.84 | 1.00 |
| PSNR | Maximum acceleration | NA | Pearson | NA | NA | NA | NA | 0.84 | 1.00 |
| PSNR | Learning | Supervised | Wilcoxon | 1.28 | 1.31 | 1 | 20 | 0.86 | 1.00 |
| PSNR | Complex input | TRUE | Wilcoxon | 1.31 | 1.31 | 9 | 12 | 0.86 | 1.00 |
| SSIM | Residual | TRUE | Wilcoxon | 1.38 | 1.52 | 12 | 14 | 0.90 | 1.00 |
| SSIM | GAN based | TRUE | Wilcoxon | 1.47 | 1.38 | 22 | 4 | 0.92 | 1.00 |
| SSIM | Learning | Supervised | Wilcoxon | 1.31 | 1.46 | 1 | 25 | 0.92 | 1.00 |
| SSIM | Cluster | GAN based | Wilcoxon | 1.47 | 1.38 | 22 | 4 | 0.92 | 1.00 |
| PSNR | Cluster | Others | Wilcoxon | 1.31 | 1.32 | 10 | 11 | 0.92 | 1.00 |
| SSIM | Architecture | U-Net like | Wilcoxon | 1.38 | 1.53 | 13 | 13 | 0.96 | 1.00 |
| SSIM | Data consistency | TRUE | Wilcoxon | 1.49 | 1.37 | 18 | 8 | 0.98 | 1.00 |
| SSIM | Parallel imaging | TRUE | Wilcoxon | 1.49 | 1.39 | 17 | 8 | 1.00 | 1.00 |

## IV. SUPPLEMENTARY FIGURES

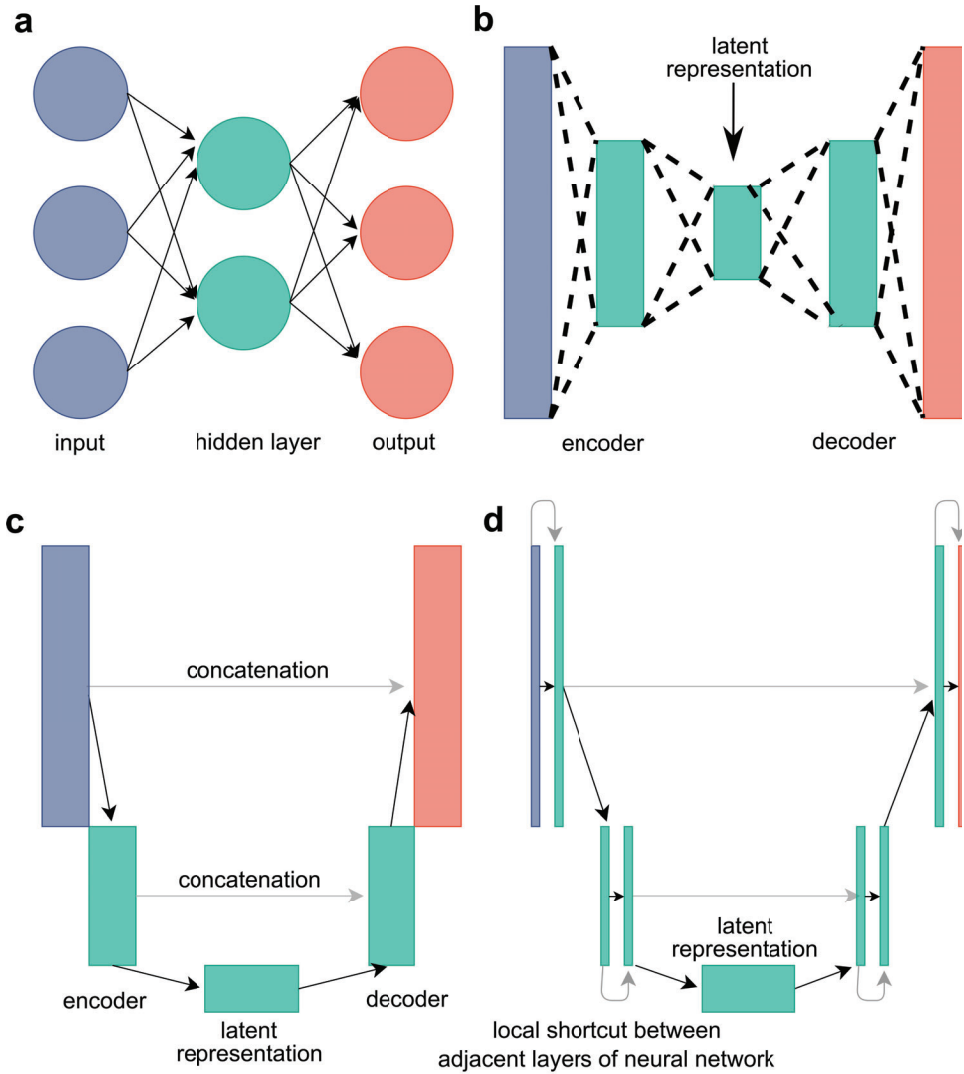

**Supplementary Figure 1.** Key neural network designs used in deep CS-MRI techniques. In each diagram, the blue nodes represent the input image, i.e., the undersampled images. The red nodes represente the reconstructed image by the network. The hidden nodes are coloured green. Connections between successive layers are shown as black arrows. Short cut connections skipping one or more layers are displayed as gray arrows. **a,** Diagram of a multilayer perceptron, one of the simplest forms of artificial neural networks. **b,** The structure of an auto-encoder. An encoder stream (left) learns a noise-free latent representation (center) of the input image and a decoder stream reconstructs this latent representation to a noise- or artifact- free image. **c,** The structure of U-Net, a successful model for image segmentation [39] and the predominant form of ANN used in 40.2% of the CS-MRI papers included in this review. It has an autoencoder-like structure with shortcut connections between the encoder and decoder streams (gray arrow). **d,** The structure of T-Net [40], [41] with a U-Net skeleton and dense local connections between adjacent layers. These 4 figures illustrate the gradual increase in complexity of neural network design with the addition of more shortcut connections.

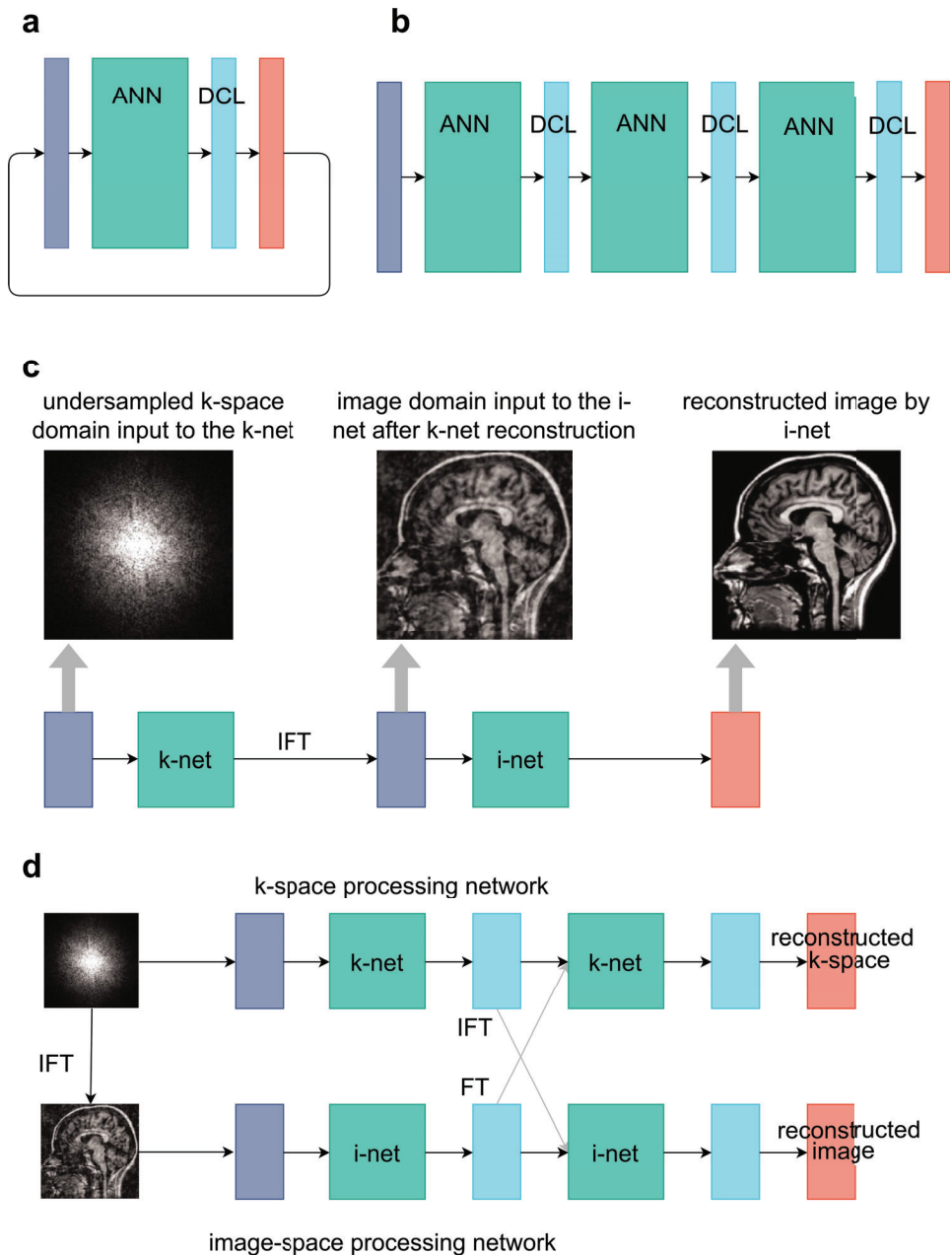

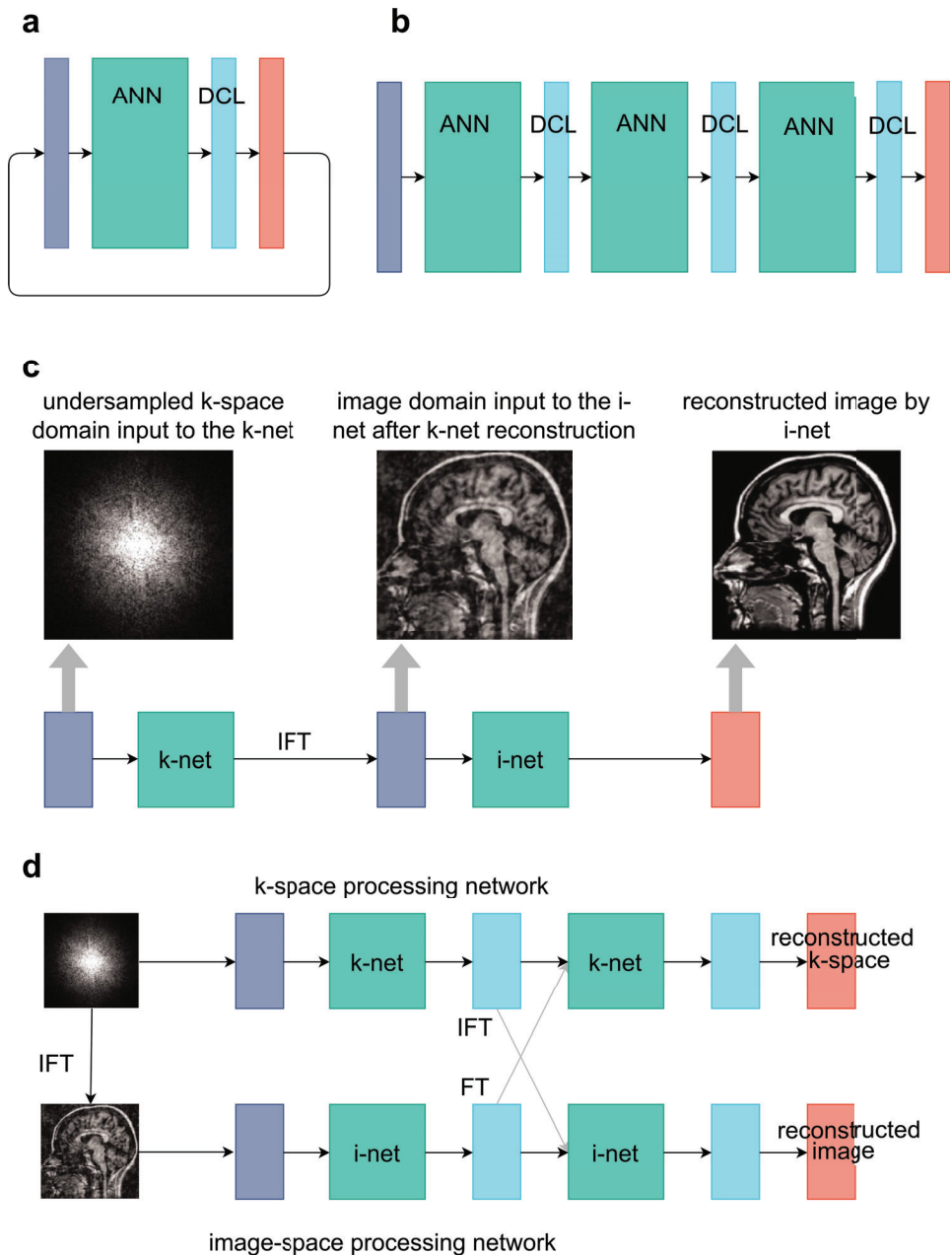

**Supplementary Figure 2.** Structure of specific deep learning models. **a**, Structure of DC-CNN. The output from the artificial neural network (ANN) is updated by a data consistency layer to enforce k-space data fidelity. These two operations are repeated to reconstruct the MR image iteratively. **b**, DC-CNN as an example of implementing unrolled optimisation in an end-to-end manner. The data consistency layer (DCL) can be seen as a computational layer following each ANN. The alternating ANN and DCL can also be viewed as a cascade of neural networks and trained in an end-to-end manner. (Figure modified from [29]). **c**, Simplified architecture of KIKI-net, the first cross domain deep CS-MRI technique. The k-net and i-net are concatenated in series to reconstruct the MR image in tandem (modified from [42]). The network is trained first to remove artifacts from the k-space domain, then from the image domain. For simplicity, only one k-net and one i-net are shown. **d**, Parallel cross domain design. Two subnetworks represent k-space and image processing streams respectively. There are shortcut connections between the hidden layers of the two subnetworks. Abbreviations: ANN, artificial neural network; DCL, data consistency layer; FT, Fourier transform to convert image space information into k-space signals; IFT, inverse Fourier transform to convert the k-space signal into image domain.

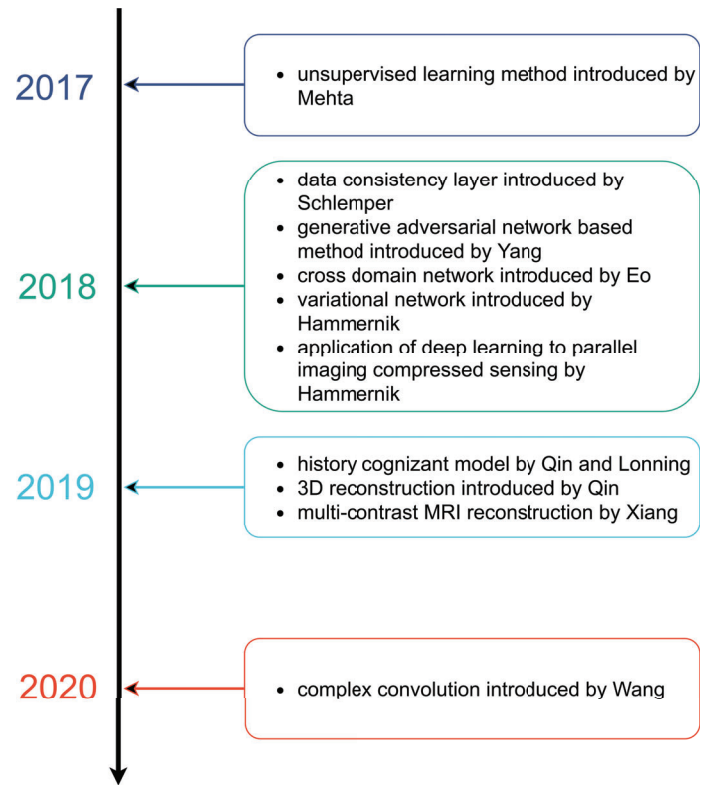

**Supplementary Figure 3.** Timeline of the key milestones in the development of deep learning-based CS-MRI techniques. A milestone feature is defined as one that is utilized in the deep learning model design by at least four subsequent studies. The year 2018 saw the highest number of significant developments.